\documentclass[usenatbib]{mnras}
\usepackage[final]{microtype}
\usepackage[T1]{fontenc}
\usepackage{ae,aecompl}
\usepackage{graphicx}
\usepackage{amsmath}
\usepackage{amssymb}	
\usepackage{subfig}
\usepackage{multirow}
\usepackage{tabularx}
\usepackage[mathscr]{euscript}
\usepackage{pifont}
\usepackage{array}
\usepackage{xcolor}
\usepackage{braket}
\usepackage{enumerate}
\usepackage{amsfonts}

\usepackage{scalefnt}
\usepackage[normalem]{ulem}
\usepackage{tikz}
\usepackage{float}
\restylefloat{table}
\newcolumntype{C}{>{$}c<{$}}
\usepackage{booktabs} 
\newcolumntype{P}[1]{>{\centering\arraybackslash}p{#1}}
\def\simless{\mathbin{\lower 3pt\hbox
{$\rlap{\raise 5pt\hbox{$\char'074$}}\mathchar"7218$}}}   
\def\simmore{\mathbin{\lower 3pt\hbox
{$\rlap{\raise 5pt\hbox{$\char'076$}}\mathchar"7218$}}}   
\newcommand{\be}{\begin{equation}}
\newcommand{\ee}{\end{equation}}
\newcommand{\MSun}{{\rm M}_{\sun}}

\newcommand{\soutPC}{\bgroup\markoverwith{\textcolor{cyan}{\rule[0.5ex]{2pt}{1pt}}}\ULon}

\title[Measuring eccentricity from GWs of LISA MBHBs]{The minimum measurable eccentricity from gravitational waves of LISA massive black hole binaries}
\author[M. Garg et al.]{Mudit Garg,$^{1}$\thanks{E-mail: mudit.garg@ics.uzh.ch} Shubhanshu Tiwari,$^2$ Andrea Derdzinski,$^1$ John G. Baker,$^3$ Sylvain Marsat,$^4$ \newauthor and Lucio Mayer$^1$\\
$^1$Center for Theoretical Astrophysics and Cosmology, Institute for Computational Science, University of Zurich,\\
Winterthurerstrasse 190, CH-8057 Z\"urich, Switzerland,\\
$^2$Physik-Institut, Universit\"at Z\"urich, Winterthurerstrasse 190, 8057 Z\"urich, Switzerland,\\
$^3$Goddard Space Flight Center, 8800 Greenbelt Rd, Greenbelt, Maryland 20771, USA,\\
$^4$Laboratoire des 2 Infinis - Toulouse (L2IT-IN2P3), \\
Universit{\'e} de Toulouse, CNRS, UPS, F-31062 Toulouse Cedex 9, France}

\date{Received / Accepted}
\pubyear{2023}
\begin{document}
\label{firstpage}
\pagerange{\pageref{firstpage}--\pageref{lastpage}}
\maketitle

\begin{abstract}
We explore the eccentricity measurement threshold of LISA for gravitational waves radiated by massive black hole binaries (MBHBs) with redshifted BH masses $M_z$ in the range $10^{4.5}$--$10^{7.5}~\MSun$ at redshift $z=1$. The eccentricity can be an important tracer of the environment where MBHBs evolve to reach the merger phase. To consider LISA's motion and apply the time delay interferometry, we employ the \textsc{lisabeta} software and produce year-long eccentric waveforms using the inspiral-only post-Newtonian model \textsc{TaylorF2Ecc}. We study the minimum measurable eccentricity ($e_{\rm min}$, defined one year before the merger) analytically by computing matches and Fisher matrices, and numerically via Bayesian inference by varying both intrinsic and extrinsic parameters. We find that $e_{\rm min}$ strongly depends on $M_z$ and weakly on mass ratio and extrinsic parameters. Match-based signal-to-noise ratio criterion suggest that LISA will be able to detect $e_{\rm min}\sim10^{-2.5}$ for lighter systems ($M_z\lesssim10^{5.5}~\MSun$) and $\sim10^{-1.5}$ for heavier MBHBs with a $90$ per cent confidence. Bayesian inference with Fisher initialization and a zero noise realization pushes this limit to $e_{\rm min}\sim10^{-2.75}$ for lower-mass binaries, assuming a $<50$ per cent relative error. Bayesian inference can recover injected eccentricities of $0.1$ and $10^{-2.75}$ for a $10^5~\MSun$ system with a $\sim10^{-2}$ per cent and a $\sim10$ per cent relative errors, respectively. Stringent Bayesian odds criterion ($\ln{\mathcal{B}}>8$) provides nearly the same inference. Both analytical and numerical methodologies provide almost consistent results for our systems of interest. LISA will launch in a decade, making this study valuable and timely for unlocking the mysteries of the MBHB evolution.
\end{abstract} 


\section{Introduction}

The Laser Interferometer Space Antenna (LISA; \citealt{AmaroSeoane2017,Barack2019}) will be one of the first space-based gravitational wave (GW) observatories that will launch in the 2030s, along with TianQin \citep{Wang2019} and Taiji \citep{Gong2021}. It will be sensitive to observed frequencies of GWs in the range of $\sim$10$^{-4}$--$10^{-1}$~Hz. The primary extragalactic sources for LISA are mergers of massive black hole binaries (MBHBs) of $10^4$--$10^8~\MSun$ and intermediate/extreme mass ratio inspirals (I/EMRIs; \citealt{Babak2017, AmaroSeoane2018b}) with primary-to-secondary BH mass ratio $q$ greater than $10^3$. LISA will be sensitive enough to detect GWs from coalescing MBHBs with $q\lesssim10.0$ up to redshift $z\sim20$ \citep{AmaroSeoane2017}. Most MBHBs will have high signal-to-noise ratios (SNRs; \citealt{AmaroSeoane2017}) in the LISA band, which will help to constrain their parameters with high accuracy.

MBHBs mainly form as by-products of galaxy mergers \citep{Begelman1980}. The process involved in shrinking the separation between MBHs from galactic scales to form a binary in the post-merger nucleus takes millions to billions of years, depending on the internal structure of the host galaxies and the relative dominance of various astrophysical processes (see, e.g. \citealt{AmaroSeoane2022}). At sub-pc scales, the interaction of the binary with gas and stars in its environment can drive the binary to the coalescence phase in the LISA band within a Hubble time \citep{Haiman2009,Milosavljevic2003}. By the time a tight binary is formed, information on its dynamical history, which reflects the nature of the properties of the host galactic nucleus, is mostly lost. However, GW waveforms from these tight systems can carry signatures of the source environment, either in the form of modifications of the vacuum waveform, from phase shifting \citep{Barausse2014,Derdzinski2019,Derdzinski2021,Toubiana2021,Sberna2022,Cardoso2022} and amplitude modulation \citep{DOrazio2020} to the injection of additional harmonics at higher frequency \citep{Zwick2022}, or via a direct relation with the binary parameters that can be extracted from the analysis of the vacuum waveform. In the latter case, the precise astrophysical environment an MBHB evolves within from pc-scales to the near-merger stage may lead to different system variables at the LISA entry for the same starting binary. 

One of the most sensitive binary parameters to the surrounding environment is the orbital eccentricity. While most studies in the literature assume that MBHBs will circularize by the time they enter the LISA band (with  entry eccentricity $e_{\rm LISA}\lesssim10^{-4}$) due to emission of GWs \citep{Peters1963, Peters1964}, some may retain non-negligible eccentricity due to evolving in a suitable dynamical environment, e.g. if MBHBs are embedded in gas \citep{Armitage2005,Sesana2005,MacFadyen2008,Cuadra2009,Roedig2011,Roedig2012,Siwek2023,Tiede2023}, in a star cluster \citep{Matsubayashi2007,Loeckmann2008,Preto2009,Sesana2010,Gualandris2022}, in a tri-axial potential \citep{Merritt2011,Khan2013}, or if they interact with a third BH \citep{Bonetti2016,Bonetti2018a,Bonetti2018b,Bonetti2019}. Hence, eccentricity can be an important tracer to probe these effects.

The eccentricity is a unique intrinsic binary parameter because it decreases rapidly as the system approaches the merger. As a result, in order to infer it from a waveform, we need to detect the GW signal many cycles before the merger. Therefore, for now, the ground-based LIGO-Virgo-KAGRA (LVK) collaboration does not include eccentricity in their analysis of the stellar-mass ($\lesssim100~\MSun$) BH binaries (SmBHBs) due to the challenges in modelling late-inspiral-merger with the presence of eccentricity and spins \citep[see, e.g.][]{RamosBuades2021}. However, LVK indeed does searches for eccentric SmBHBs using un-modelled methods \citep{Abbott2019, RamosBuades2020}. Given that we will observe GWs in the early inspiral phase in the LISA band for most MBHBs, ignoring eccentricity could lead to mismodelling of the GW waveform. Most of the focus on eccentricity detection in the LISA frequency band has been in the context of multi-band SmBHBs sources \citep{Nishizawa2016,Nishizawa2017,Klein2022}, with some attention on EMRIs. Multi-band sources are seen in the LISA band a few years before they merge in the LVK frequency band of $\sim10$--$10^4$ Hz \citep{Sesana2016,Vitale2016}. The detection of eccentricity is proposed as a way to distinguish whether SmBHBs are formed in the field or via dynamical interaction such as in globular clusters, nuclear clusters, or galactic nuclei  \citep{Nishizawa2016,Breivik2016,Samsing2018,DOrazio2018,Gondan2018b,RomeroShaw2019,RomeroShaw2020,Zevin2021,RomeroShaw2022}. Also, eccentricity can help in breaking parameter degeneracies by inducing higher harmonics \citep{Mikoczi2012,Yang2022,Xuan2022} and it can improve parameter estimation accuracy \citep{Sun2015,Vitale2016,Gondan2018b,Gondan2019, Gupta2020}. EMRIs are mostly expected to have a significant entry eccentricity in the LISA band, ranging from $e_{\rm LISA}\gtrsim0.1$--$0.8$ \citep{HopmanAlexander2005,AmaroSeoane2018a}, which can be measured to high accuracy, barring data analysis challenges \citep{Babak2017,Berry2019,Chua2022}.

This work considers eccentric binaries in vacuum of two near-coalescence non-spinning MBHs. We are interested in determining the minimum eccentricity that can be confidently measured by LISA one year before merger for a given MBHB source at $z=1$. Our analysis attempts to be as realistic as possible in the data analysis which will be employed for LISA once the mission is operational, i.e. we take into account the full LISA motion in its orbit around the Sun, generate high-order post-Newtonian (PN) waveforms, employ the time delay interferometry (TDI) technique to cancel the detector's laser noise, and finally perform Bayesian inference to recover injected parameters.

The measurability of eccentricity in the MBHB’s GW waveform is a novel investigation. It is an important study because, similar to multi-band sources, residual eccentricities can be a signature of the environment in which MBHBs have evolved. For instance, recent high-resolution hydrodynamical simulations by \citet{Zrake2021} show that for equal-mass binaries hardening in prograde circumbinary gas discs, we expect an eccentricity of $\sim10^{-3}$ one year before coalescence. The eccentricity evolution in the late stages of hardening by a prograde accretion disc is further supported by \citet{DOrazio2021} and \citet{Siwek2023}. Moreover, \citet{Tiede2023} show that we should expect even higher eccentricity in the LISA band if the circumbinary disc is retrograde instead of prograde. Therefore, eccentricity detection by LISA could be a tracer of gas interaction. Simulations of MBH binary evolution starting from realistic galaxy mergers \citep{Capelo2015}, in which three-body encounters with stars dominate the orbital decay at sub-pc separations, show that the eccentricity always increases above the value that it has when the hardening phase begins, reaching values as large as 0.9 \citep{Khan2018}. The residual value of eccentricity around $50$--$100$ Schwarzschild radii (about one year before merger), when circularization via GW emission has already started to act, is yet to be determined. However, recently \citet{Gualandris2022} studied the evolution of eccentricity through
the stellar hardening phase and into the GW radiation regime, finding that the residual value of the eccentricity at about $50$ Schwarzschild radii for a $4\times10^6~\MSun$ MBHB ranges from below $10^{-4}$ to nearly $10^{-3}$ (as suggested by Elisa Bortolas in further communication). Interestingly, the specific eccentricity here mainly depends upon the parameters at large scale and positively correlates with the initial eccentricity of the merging galaxies. Also, the lowest possible eccentricity detectable by LISA for a given MBHB will tell us whether its neglect during parameter estimation will lead to biases and degeneracies. The consensus for entry eccentricity in the LISA band for MBHBs is $\lesssim10^{-4}$, which justifies the circular assumption, but for $e_{\rm LISA}>10^{-4}$, it would be crucial to consider eccentricity during match filtering and when constraining binary variables \citep{Porter2010}. 

The paper is structured as follows. In Section~\ref{Sec:waveform}, we describe our waveform model and systems of interest. Section~\ref{Sec:analytical} studies analytical constraints on eccentricity measurement using matches and Fisher formalism. In Section~\ref{Sec:Bayesian}, we detail our Bayesian setup to find the minimum measurable eccentricity. We discuss the findings in Section~\ref{Sec:discussion} and summarize the key takeaways of this work in Section~\ref{Sec:conclusion}.

\section{Waveform generation, System parameters, LISA response, and time delay interferometry}\label{Sec:waveform}

MBHBs are one of the most promising sources for LISA as they are expected to be the loudest events and will spend a significant amount of time (up to a few years) in LISA's frequency band before merging. Most of the time MBHBs spend in the LISA band is in the long-inspiral phase where eccentricity (e) can still be non-negligible. The inspiral part of the GW waveforms from eccentric BHB mergers has been developed within the PN formalism both in time and frequency domains \citep{Damour2004, Mishra2016}. The time-domain PN waveforms have the advantage of having a larger region of validity in eccentricity, and they can probe eccentricities up to $\approx 0.8$, but they are slow to generate \citep{Tanay2016,Tanay2019}. The frequency-domain waveforms are much faster to compute but are limited to the low-eccentricity approximation. For LISA data analysis, it is imperative to have fast waveform computation as the evolution of the BHB occurs over a large time-frequency volume. There exist a wide range of frequency-domain eccentric BHB waveforms, namely \textsc{TaylorF2Ecc} \citep{Moore2016}, \textsc{EccentricFD} \citep{Huerta2014}, and \textsc{EFPE} \citep{Klein2021}, among others. For this study, we have employed the \textsc{TaylorF2Ecc} inspiral-only waveform model with circular phasing accurate up to $3.5$PN order taken from another inspiral-only model \textsc{TaylorF2} \citep{Buonanno2009} and eccentricity corrections to phasing up to 3PN and $\mathcal{O}(e^2)$, making it valid for $e\lesssim0.1$. However, this model does not give a prescription for spinning BHs. We choose \textsc{TaylorF2Ecc} as our fiducial model as astrophysically we mostly do not expect higher eccentricities, as mentioned in the introduction.

The parameter space we consider for MBHBs spans the range of total redshifted masses $M_z$ between $10^{4.5}$--$10^{7.5}~\MSun$, mass ratios\footnote{q=1 system gives leads to Fisher initialization problems in Bayesian inference, hence we choose $q=1.2$ as a representative value.  } $q\in[1.2,8]$, and the initial eccentricity one year before the merger $(e_{1{\rm yr}})$ between $10^{-3.5}$--$0.1$. We have not considered the individual spins of the component BHs for this work. Unless otherwise stated, we always quote the values at the detector frame (L-frame). This leaves us with three intrinsic parameters\footnote{While there are other eccentricity-related binary parameters, we only focus on the magnitude of eccentricity.} (first three rows of Table~\ref{table:parameters}) and six extrinsic parameters (last six rows of Table~\ref{table:parameters}). We employ the cosmological parameters from the \citet{Planck2020} survey to compute the luminosity distance from redshift: Hubble constant $H_0=67.77~\text{km}~\text{s}^{-1}~\text{Mpc}^{-1}$, matter density parameter $\Omega_{\rm m}=0.30966$, and dark-energy density parameter $\Omega_{\Lambda}=0.69034$.

\begin{table}
\centering
    \begin{tabular}{|p{0.6\linewidth}|c|}
        \hline
        {\bf Parameter} & {\bf Units}\\
        \hline
        \hline
        Total redshifted mass $M_z$ & $\MSun$\\
        \hline
        Mass ratio $q$ & Dimensionless\\
        \hline
        Eccentricity one year before coalescence $e_{1{\rm yr}}$  & Dimensionless\\
        \hline
        Luminosity distance $D_{\rm L}$ & Mpc\\
        \hline
        Phase at coalescence $\phi$ & Radian\\ 
        \hline
        Inclination $\imath$ & Radian \\ 
        \hline
        Ecliptic latitude $\lambda$  & Radian \\ 
        \hline
        Ecliptic longitude $\beta$ & Radian \\ 
        \hline
        Initial polarization angle $\psi$ & Radian \\ 
        \hline
    \end{tabular}
\caption{Source parameters in the L-frame.}
\label{table:parameters}
\end{table}
 
We generate eccentric waveforms $\tilde{h}_{\rm ecc}(f)$ for our systems of interest using the \textsc{TaylorF2Ecc} template over these parameter grids to optimally cover the intrinsic parameter space:
\begin{align}\label{eq:parameters}
    M_z\in&\{10^{4.5},10^5,10^{5.5},10^6,10^{6.5},10^{7},10^{7.5}\}~\MSun,\\
    q\in&\{1.2,2.0,4.0,8.0\},\nonumber\\
    e_{1{\rm yr}}\in&\{10^{-3.5},10^{-3.25},10^{-3},10^{-2.75},10^{-2.5},10^{-2.25},\nonumber\\
    &10^{-2},10^{-1.75},10^{-1.5},10^{-1.25},0.1\}\nonumber.
\end{align}
Additionally, we also generate quasicircular ($e_{1{\rm yr}}=0$) waveforms ($\tilde{h}_{\rm cir}$). For extrinsic parameters, our fiducial values are z=1 which corresponds to $D_{\rm L}=6791.3$~Mpc, and the angles are all set to $0.5$ radians. We choose these parameters such that MBHBs spend at least one year in the LISA band before coalescing.

In this work, we only work with Newtonian amplitudes and study binaries until they reach their innermost stable circular orbit (ISCO), i.e. at binary separation $r_{\rm ISCO}\equiv 3r_{\rm s}$, where $r_{\rm s} \equiv2GM_z/c^2$ is the Schwarzschild radius of the total mass BH,\footnote{$G$ is the gravitational constant and $c$ is the speed of light in vacuum.} at which point binaries are expected to circularize.\footnote{We perform circularization test of \textsc{TaylorF2Ecc} in Appendix~\ref{AppA}.} We find the starting frequency ($f_{1{\rm yr}}$) such that the system reaches the ISCO at frequency $f_{\rm ISCO}$ in exactly one year by using the Peters' time-scale \citep{Peters1964}. The reason why we have chosen $f_{1{\rm yr}}$ as our reference frequency to initiate the signal and not some fixed frequency as usually employed by the LVK collaboration is that MBHB masses can vary by up to three orders of magnitude, making any fixed frequency sufficient for some systems and too short or long for others.

\begin{figure}
    \centering
    \includegraphics[width=\linewidth]{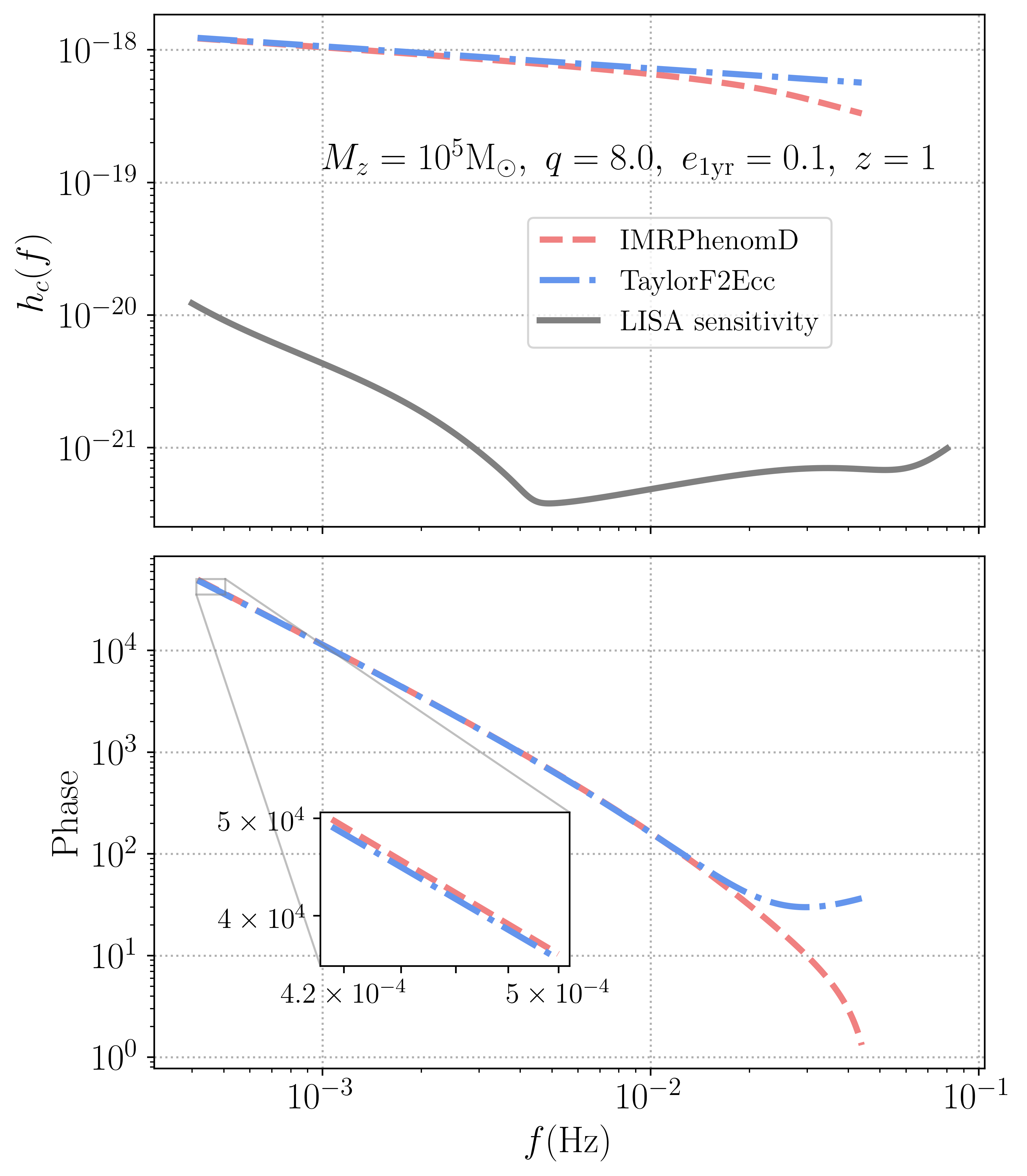}
    \caption{Characteristic strain $h_{\rm c}$ compared to the LISA noise (solid gray; in the top panel) and accumulated phase (in the bottom panel) for an $M_z=10^5~\MSun$, $q=8.0$, and $e_{1{\rm yr}}=0.1$ binary at $z=1$ for waveform templates \textsc{IMRPhenomD} (dashed red) and \textsc{TaylorF2Ecc} (dot-dashed blue) between $f_{1{\rm yr}}$ and $f_{\rm ISCO}$. In the bottom panel we also enlarge the initial phase to show the difference between quasicircular and eccentric phasings.}
    \label{fig:waveforms}
\end{figure}

In Fig.~\ref{fig:waveforms}, we show the characteristic strain $h_{\rm c}(f)\equiv 2f\tilde{h}(f)$, a visual aid to represent how signal adds up in the detector, the LISA noise curve $S_n(f)$ including a confusion noise due to galactic binaries taken from \citet{Marsat2021}, and the accumulated phase for an $M_z=10^5~\MSun$, $q=8.0$, and $e_{1{\rm yr}}=0.1$ system at $z=1$ for \textsc{TaylorF2Ecc} and the quasicircular inspiral-merger-ringdown waveform model \textsc{IMRPhenomD} \citep{Husa2016,Khan2016}. Since the inspiral part of the \textsc{IMRPhenomD} comes from the \textsc{TaylorF2} template, phasings are almost identical until the system is close to the ISCO. The initial phase difference is due to non-zero eccentricity in \textsc{TaylorF2Ecc}, and later deviations come from the merger part of the \textsc{IMRPhenomD}, which are beyond the scope of the inspiral-only model \textsc{TaylorF2Ecc}.

To account for LISA motion and to project the waveform into TDI channels, namely A, E, and T, we modify the \textsc{lisabeta} software (see \citealt{Marsat2021} for details and subsequent notations) by including support for \textsc{TaylorF2Ecc}. Therefore, a waveform $\tilde{h}(f)$ will have strain projections $\tilde{h}_{\rm A,E,T}(f)$ and noise power spectral densities $S_n^{\rm A,E,T}$ corresponding to A, E, and T channels, respectively.

\begin{figure}
    \centering
    \includegraphics[width=\linewidth]{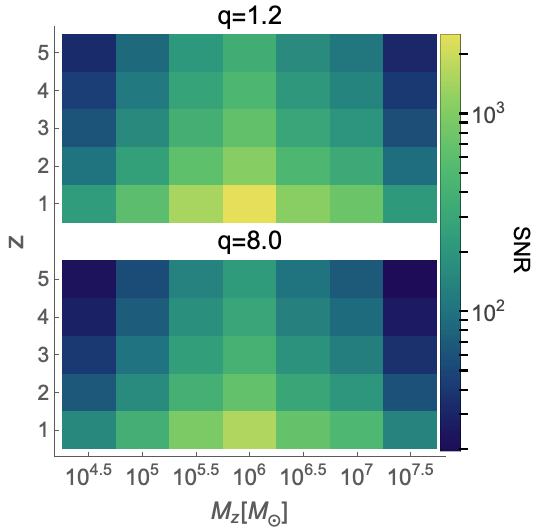}
    \caption{SNR for our systems of interest for two limiting mass ratios of $q=1.2$ (top panel) and $q=8.0$ (bottom panel). In both panels, we vary $M_z$ from $10^{4.5}$ to $10^{7.5}~\MSun$ and $z$ from $1$ to $5$, and set rest of the parameters to our fiducial values. These SNRs take into account LISA motion and are calculated by summing over three TDI channels A, E, and T. The low eccentricities we consider here do not affect the SNR significantly.}
    \label{fig:SNR}
\end{figure}

Now, we can write down the standard inner-product between two waveforms as
\begin{equation}
    (a|b)=4\sum_{\rm A,E,T}{\rm Re}\int^{f_{\rm ISCO}}_{f_{1{\rm yr}}}{\rm d}f\frac{\tilde{a}(f)\tilde{b}^*(f)}{S_{\rm n}(f)},
\end{equation}
\noindent where the pre-factor 4 comes from the one-sided spectral noise density normalization. Hence, the SNR of the signal is $\sqrt{(h|h)}$. In Fig.~\ref{fig:SNR}, we show the dependence of the SNR at $z=1$ as a function of total mass and mass ratio for our parameter space in Eq.~\eqref{eq:parameters}. As expected, the SNR is higher for near-equal mass ratios than the unequal ones and decreases as the redshift increases. Furthermore, the SNR peaks at middle-range masses of $\sim10^6~\MSun$, known as golden LISA sources.

In the following two sections, we find the minimum eccentricity and errors on its recovery in the LISA data stream for a given source. First, we approach this task analytically by using a match between waveforms and computing Fisher information matrices. We then perform Bayesian inference to numerically determine the posteriors.

\section{Analytical measurability of eccentricity}\label{Sec:analytical}

We first present a simple and commonly used estimate for the distinguishability of eccentric from quasicircular binaries in LISA using a match-based SNR criterion defined in Eq.~\eqref{eq:Optmatch}. Furthermore, we employ the Fisher formalism to estimate how well-constrained eccentricity will be for these sources. These computations provide a theoretical benchmark that can be compared with Bayesian inference presented later.

\subsection{Optimal match}\label{Subsec:match}

\begin{figure}
    \centering
    \includegraphics[width=\linewidth]{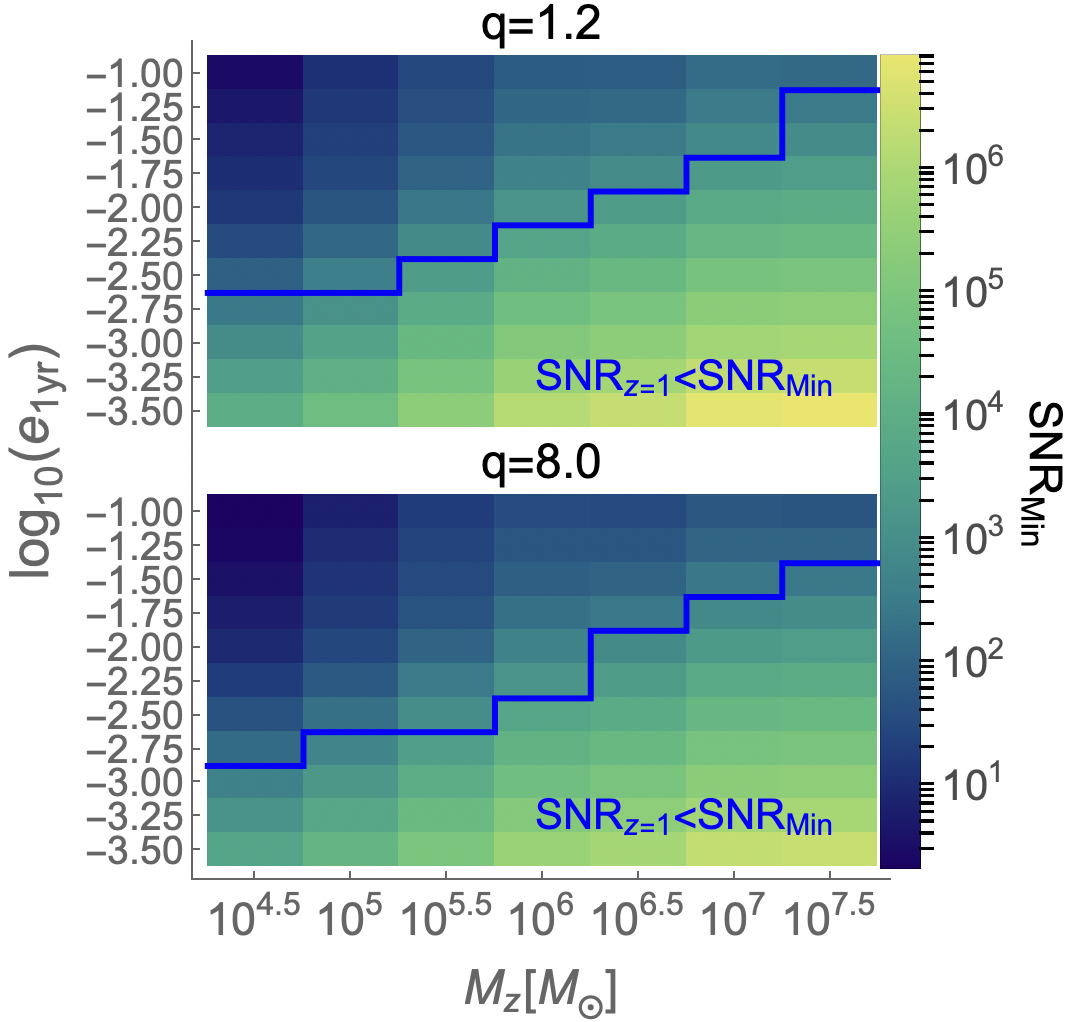}
    \caption{${\rm SNR}_{\rm min}$ required to distinguish between quasicircular and eccentric waveforms for our parameter space. In the top panel, we fix $q=1.2$, and vary $M_z$ from $10^{4.5}$ to $10^{7.5}~\MSun$ and $e_{1{\rm yr}}$ from $10^{-3.5}$ to $10^{-1}$. In the bottom panel, we keep the mass ratio $q=8.0$ constant, and vary $M_z$ and $e_{1{\rm yr}}$ as in the top panel. Both panels have a blue line showing the boundary of the LISA non-detectability region at $z=1$ (${\rm SNR}_{z=1}<{\rm SNR}_{\rm min}$).}
    \label{fig:Optmatch}
\end{figure}

We compute matches between $h_{\rm ecc}$ and $h_{\rm cir}$ waveforms with the same $M_z$ and $q$, and find the minimum SNR (${\rm SNR}_{\rm min}$) for which LISA can distinguish between these waveforms with more than $90$ per cent confidence. To compute ${\rm SNR}_{\rm min}$, we use the criterion from \citet{Baird2013}:
\begin{align}\label{eq:Optmatch1}
    {\rm SNR}_{\rm min}^2= \frac12\frac{\chi_k^2(1-p)}{(1-\mathscr{M}(h_{\rm ecc},h_{\rm cir}))},
\end{align}
\noindent where $\chi_k^2(1-p)$ is the $\chi^2$ probability distribution function, $1-p$ is the significance level, $k$ is the number of free binary parameters, and $\mathscr{M}(h_{\rm ecc},h_{\rm cir})$ is a match between ${h}_{\rm cir}$ and ${h}_{\rm ecc}$:
\begin{equation}\label{eq:match}
    \mathscr{M}(h_{\rm cir},h_{\rm ecc})=\underset{\Delta\phi}{\rm max}\frac{(h_{\rm cir}|h_{\rm ecc})}{\sqrt{(h_{\rm cir}|h_{\rm cir})}\sqrt{(h_{\rm ecc}|h_{\rm ecc})}},
\end{equation}
\noindent which is maximized over phase shifts $\Delta\phi$. In our case, we have $p=0.9$ and $k=3$ as we vary only three binary parameters -- $M_z$, $q$, and $e_{1{\rm yr}}$. This transforms Eq.~\eqref{eq:Optmatch1} into
\begin{align}\label{eq:Optmatch}
    {\rm SNR}_{\rm min}^2= \frac{3.12}{(1-\mathscr{M}(h_{\rm circ},h_{\rm ecc}))}.
\end{align}
If the event's SNR is less than ${\rm SNR}_{\rm min}$ then one cannot differentiate between quasicircular and eccentric binaries, which in turn provides a constraint on the minimum detectable eccentricity ($e_{\rm min}$) assuming the rest of the binary parameters are known. In Fig.~\ref{fig:Optmatch}, we show ${\rm SNR}_{\rm min}$ for our systems of interest and compare it with the event's SNR at our fiducial $z=1$ (${\rm SNR}_{z=1}$). It illustrates that $e_{\rm min}\sim10^{-2.5}$ for lower-mass MBHBs ($M_z\lesssim10^{5.5}~\MSun$) and $\sim10^{-1.5}$ for higher-mass systems. The strong dependence on the total mass can be attributed to the fact that even though our considered binaries spend one year in the LISA band, $f_{1{\rm yr}}$ for heavier systems is much lower than for the lighter binaries. This implies that the inspiral part of the signal, where eccentricity is dominant, will fall within the low sensitivity region of LISA, leading to systematically worse constraints for heavier systems. Moreover, the weak dependence on the mass ratio can be explained from our definition of $e_{1{\rm yr}}$. Unsurprisingly, higher eccentricities are easily distinguishable from lower ones.

One can use the SNR estimates in Fig.~\ref{fig:SNR} for any MBHBs at higher redshift in the LISA band and use Fig.~\ref{fig:Optmatch} to assess whether eccentricity will be detectable for the given system since ${\rm SNR}_{\rm min}$ is computed in the L-frame. In the next section, we find the expected error bars on the recovery of injected eccentricity using the Fisher formalism.

\subsection{Fisher matrix}\label{Subsec:Fisher}

A standard parameter estimation technique in the LISA community is to compute a Fisher matrix \citep{Vallisneri2008}, which tells us how well we can constrain a certain parameter assuming a Gaussian noise and high SNR. We can define the Fisher matrix as
\begin{equation}
    \Gamma_{\rm ab}=\left(\partial_{\rm a} h|\partial_{\rm b} h\right),
\end{equation}
\noindent where $\partial_{\rm a} h\equiv \partial h/\partial\theta_{\rm a}$ is the partial derivative of a waveform $h$ with respect to a parameter $\theta_{\rm a}$.

The inverse of the Fisher matrix is the variance-covariance matrix, whose diagonal elements are variances ($\sigma^2$) for each of the injected parameters. The square-root of a variance provides the standard deviation ($\sigma$), which tell us the error estimate on a given parameter.

\begin{figure}
    \centering
    \includegraphics[width=0.5\textwidth]{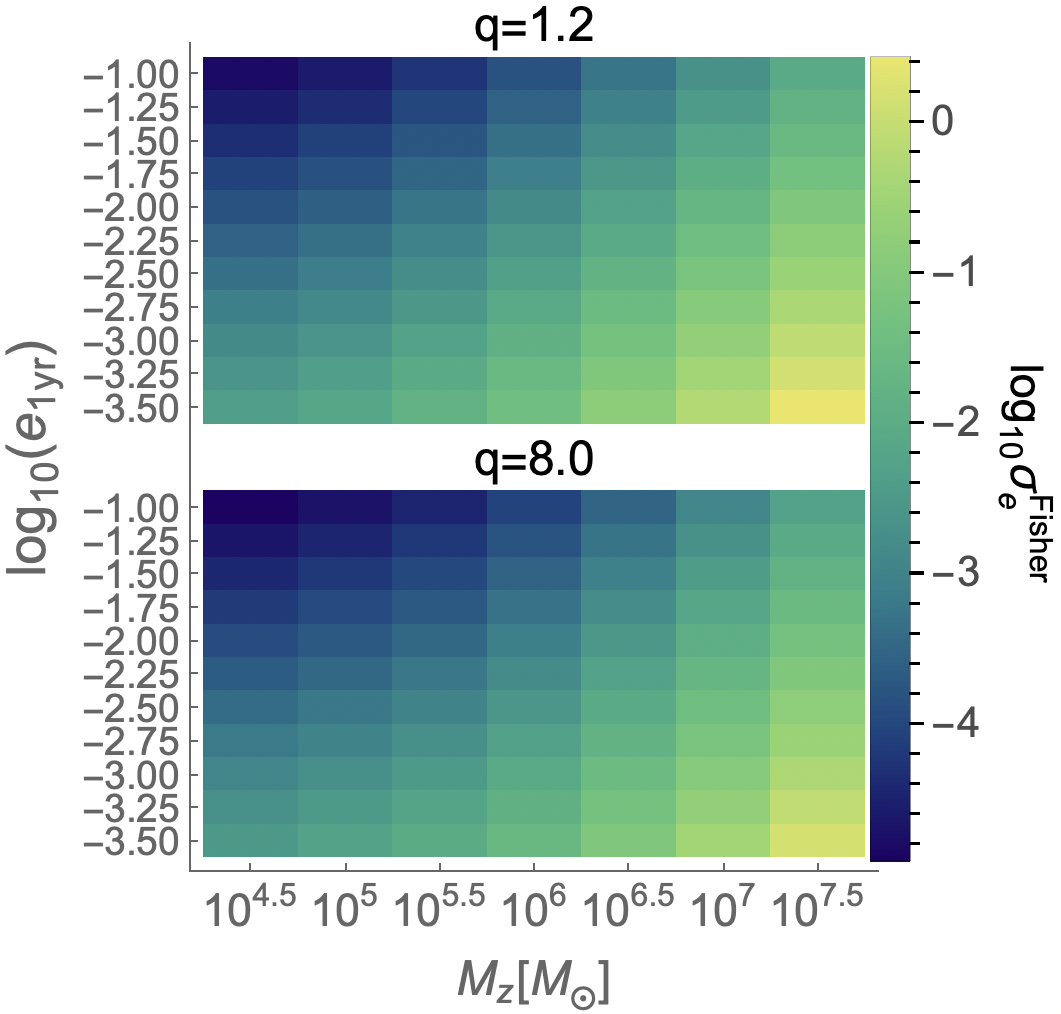}
    \caption{For the same binary parameters as in Fig.~\ref{fig:Optmatch}, error estimates by Fisher formalism ($\sigma^{\rm Fisher}_e$) on eccentricities of our considered binaries.}
    \label{fig:Fisher}
\end{figure}

We again only vary intrinsic parameters: $M_z$, $q$, and $e_{1{\rm yr}}$, and show in Fig.~\ref{fig:Fisher} the Fisher-based error estimate on eccentricity ($\sigma^{\rm Fisher}_{\rm e}$) for our parameters of interest in Eq.~\eqref{eq:parameters}. Errors mainly vary with total mass and less significantly with mass ratio, due to the same reasons as explained for the match results in Section~\ref{Subsec:match}. Fig.~\ref{fig:Fisher} suggests that for lighter systems, higher eccentricities are constrained to error (relative error $\equiv100\times\sigma^{\rm Fisher}_{\rm e}/e$) $\sim10^{-4}$ ($\sim0.1$ per cent)  whereas for lower $e_{1{\rm yr}}$ we find $\sigma^{\rm Fisher}_{\rm e}\sim10^{-3}$ ($\sim1000$ per cent). For heavier binaries, errors are $\sim10^{-3}$ ($\sim1$ per cent) for higher eccentricities and $\sim10^{-1}$ ( $10^5$ per cent) for lower $e_{1{\rm yr}}$. This suggests that lower eccentricities are completely unconstrained.

One can always scale these errors ($\sim1/$SNR) at a further luminosity distance by using SNR values in Fig.~\ref{fig:SNR} to get rough estimates. In the next section, we perform Bayesian inference to find error estimates on eccentricity recovery and the minimum measurable eccentricity.

\section{Measurability of eccentricity using Bayesian inference}\label{Sec:Bayesian}

The main goal of Bayesian inference is to construct posterior distributions $p(\theta|d)$ for the parameter space $\theta$ to fit the observed data $d$ (see, e.g. \citealt{Thrane2019}). $p(\theta|d)$ represents the probability distribution function of $\theta$ given the data $d$ and it is normalized such that $\int{\rm d}\theta~p(\theta|d)=1$. To compute the posterior, we use Bayes theorem,
\begin{equation}
    p(\theta|d)=\frac{\mathcal{L}(d|\theta)\pi(\theta)}{Z},
\end{equation}
\noindent where $\mathcal{L}(d|\theta)$ is the likelihood function of the data $d$ given the parameters $\theta$, $\pi(\theta)$ is the prior on $\theta$, and $Z\equiv\int{\rm d}\theta \mathcal{L}(d|\theta)\pi(\theta)$ is the evidence. Since we are not selecting between different models, we can treat $Z$ as a normalization constant. Also, we only consider uniform (flat) priors for all parameters.

For our stationary Gaussian noise $S_n^{A,E,T}$, we can write down the log-likelihood with a zero-noise realization summed over A, E, and T channels as \citep{Marsat2021}:
\begin{equation}\label{eq:likelihoof}
    \ln\mathcal{L}\propto\sum_{A,E,T}({h}-{h}_{\rm inj}|{h}-{h}_{\rm inj}),
\end{equation}
\noindent where $\tilde{h}$ is the template signal, and $\tilde{h}_{\rm inj}$ is the simulated injected signal. The zero-noise realization accelerates the likelihood computation, improves upon the Fisher results by providing the shape of posteriors, and helps understand parameter degeneracies and detectability of certain effects (here eccentricity). 

For sampling, we use the parallel tempering Markov-chain Monte Carlo (MCMC) code \textsc{ptmcmc}.\footnote{https://github.com/JohnGBaker/ptmcmc} To further speed up the likelihood computation, we draw random samples from a multivariate Gaussian with the mean given by the injected parameters and standard deviations provided by the Fisher formalism\footnote{Using a wider prior does not affect results as shown in Appendix~\ref{AppD}} in Section~\ref{Subsec:Fisher}.

\begin{figure}
    \centering
    \includegraphics[width=0.5\textwidth]{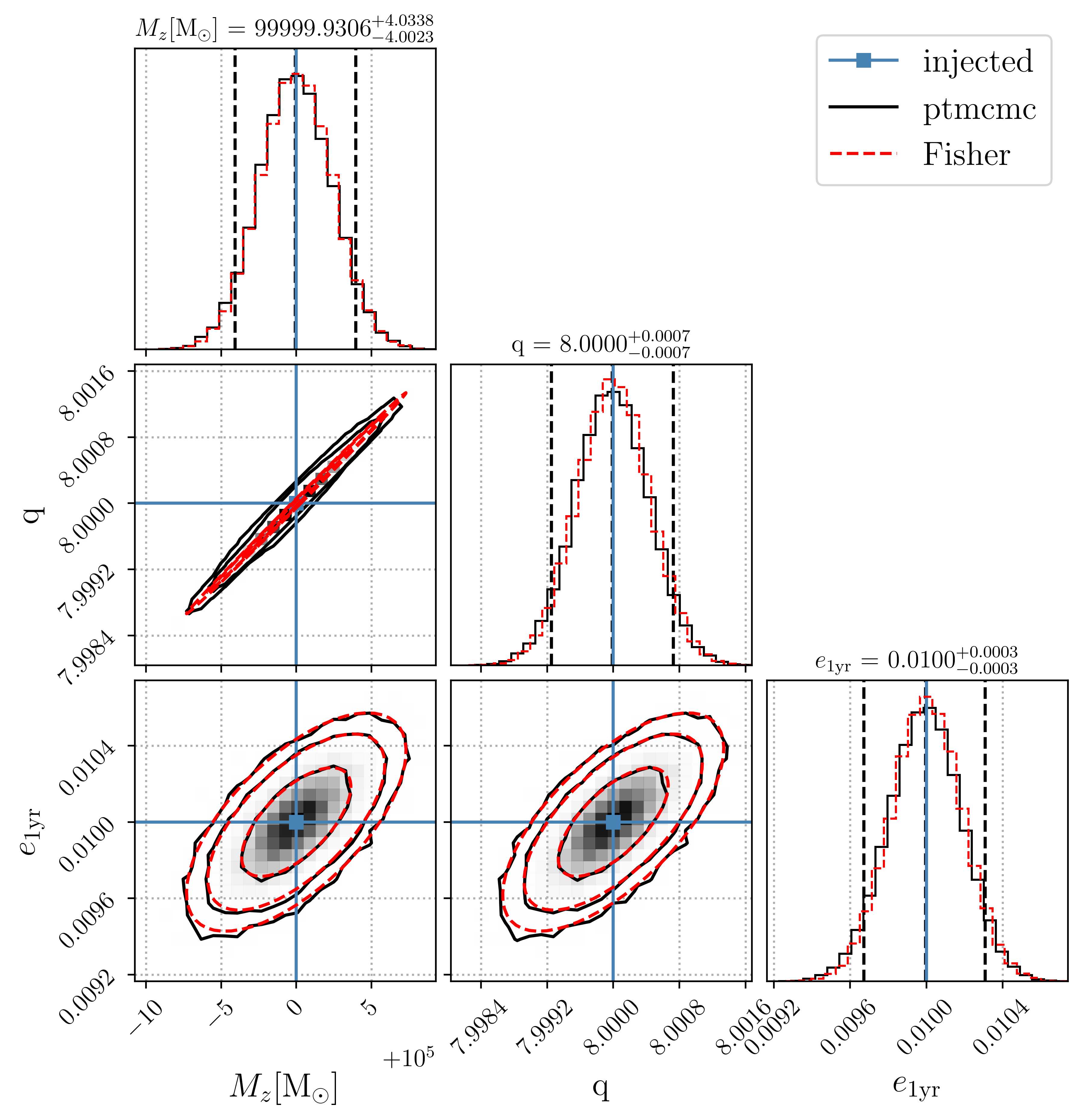}
    \caption{Posterior distributions (solid black) for an injected binary with $M_z=10^5~\MSun$, $q=8.0$, and $e_{1{\rm yr}}=0.01$. The extrinsic parameters are fixed to our fiducial values. The two extreme vertical dashed lines constrain the $90$ per cent credible interval, whereas the middle dash line represents the median of the distribution. The blue lines mark the injected values, whereas the contours in two-dimensional posteriors indicate $68$, $95$, and $99$ per cent credible intervals. We also indicate the Fisher results (dashed red) for comparison.}
    \label{fig:MCMC}
\end{figure}

We primarily sample only the intrinsic parameters and set a high-frequency cutoff for the data at $f_{\rm ISCO}$ of the injected binary.\footnote{Using an earlier cutoff than the ISCO does not significantly affect the posteriors, as shown in Appendix~\ref{AppC}.} We show the posteriors for $M_z$, $q$, and $e_{1{\rm yr}}$ in Fig.~\ref{fig:MCMC} for injected binary parameters $10^5~\MSun$, $8.0$, and $0.01$. All parameters are well recovered, with the injected values being extremely close to the median of their respective posterior. The chirp mass parameter $\mathcal{M}\equiv M(q/(1+q)^2)^{3/5}$ is even better constrained to $24932.3377^{+0.0808}_{-0.0815}~\MSun$ as expected \citep[see, e.g.][]{Cutler1994} with the injected value being $24932.3365~\MSun$. Moreover, Bayesian errors are similar to the errors provided by the Fisher formalism, as expected due to the high SNR and a zero-noise realization.

\begin{figure}
    \centering
    \includegraphics[width=0.5\textwidth]{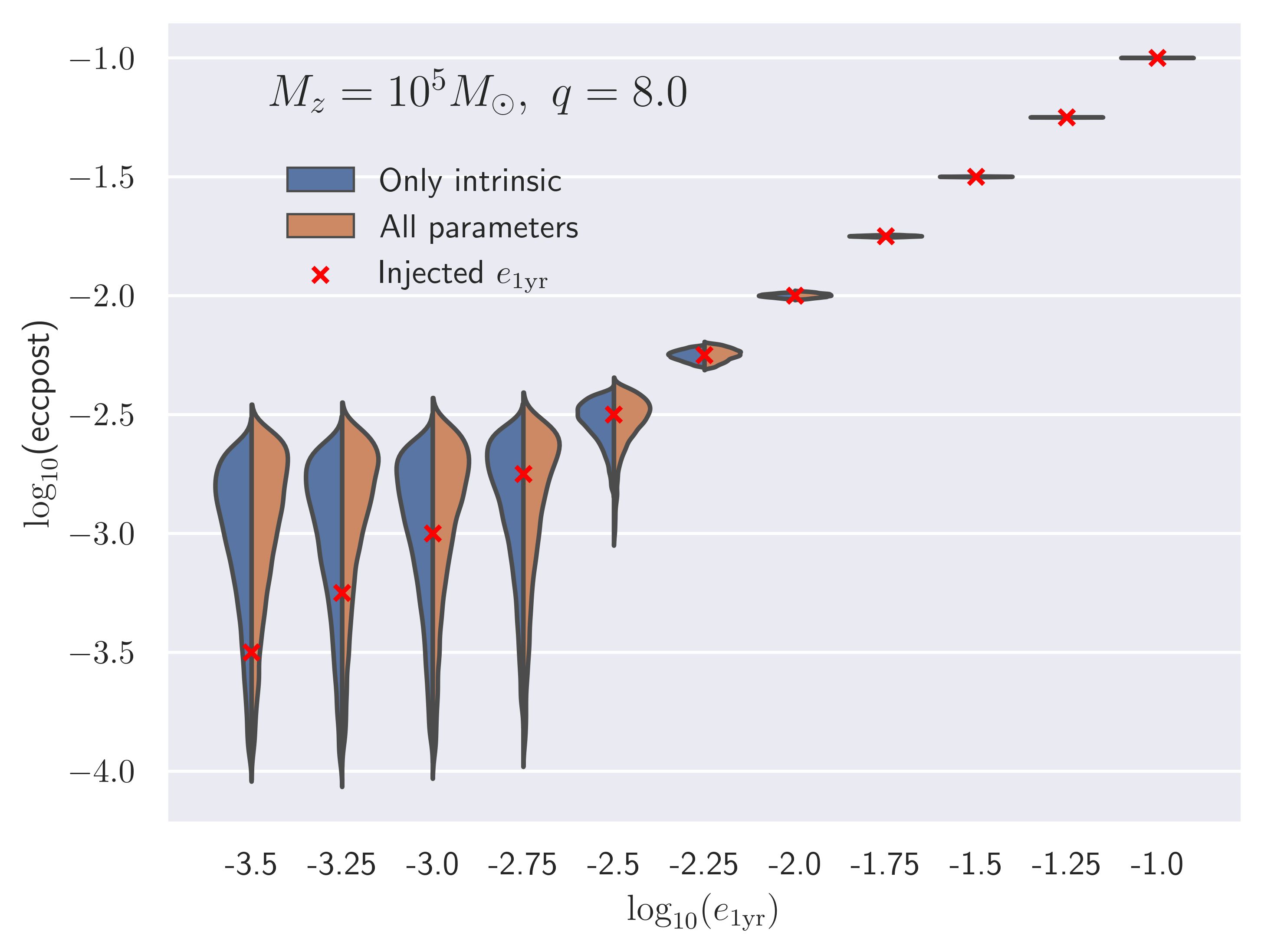}
    \caption{Posterior distributions (eccpost) for the eccentricity corresponding to each injected $e_{1{\rm yr}}$ for binaries with fixed $M_z=10^5~\MSun$ and $q=8.0$. The posteriors are constrained to the $90$ per cent credible interval and are shown in blue (left) if we only sample the intrinsic parameters and in orange (right) if we vary all parameters. The injected values are marked with a red cross. }
    \label{fig:Violin_ecc}
\end{figure}

We also study the effect of including extrinsic parameters\footnote{We show the full posteriors in Fig.~\ref{fig:MCMC_all}.} (also given in Table \ref{table:parameters}) on the measurability of the eccentricity in Fig.~\ref{fig:Violin_ecc}. Here, we show the comparison of the posteriors of $e_{1{\rm yr}}$ in Eq.~\eqref{eq:parameters} for fixed $M_z=10^5~\MSun$ and $q=8.0$ between the cases when sampling only intrinsic parameters and when sampling over all parameters in Table~\ref{table:parameters}. Adding extrinsic parameters results in a slight broadening of eccentricity posteriors and a narrow shift in the peak. This is anticipated due to the increase in degrees of freedom, which do not contribute to the measurement of eccentricity.\footnote{Eccentricity is not expected to be correlated to the extrinsic parameters.} Unsurprisingly, the higher the eccentricity, the better the recovery of the injected value, i.e. the injected value is extremely close to the peak of the posterior. 

To measure how well injected eccentricities are recovered in our Bayesian inference, we introduce a Bayesian relative error metric in terms of the injected eccentricity $e_{\rm inj}$ and the standard deviation of the corresponding eccentricity posterior $\sigma^{\rm MCMC}_e$:
\begin{equation}\label{eq:mineccMCMC}
    \sigma^{\rm MCMC}_{e,\rm rel}[\%]=100\times\frac{\sigma^{\rm MCMC}_e}{e_{\rm inj}}.
\end{equation}

To survey the parameter space widely with Bayesian inference, we have conducted a total of $7\times4\times8$ runs by sampling over only intrinsic parameters for seven values of the total mass, four values of the mass ratio, and eight values of the eccentricity given in Eq.~\eqref{eq:parameters}. We present only the runs for the intrinsic parameters here, as we have shown that including extrinsic parameters does not affect the results significantly. 

\begin{figure}
    \centering
    \includegraphics[width=0.5\textwidth]{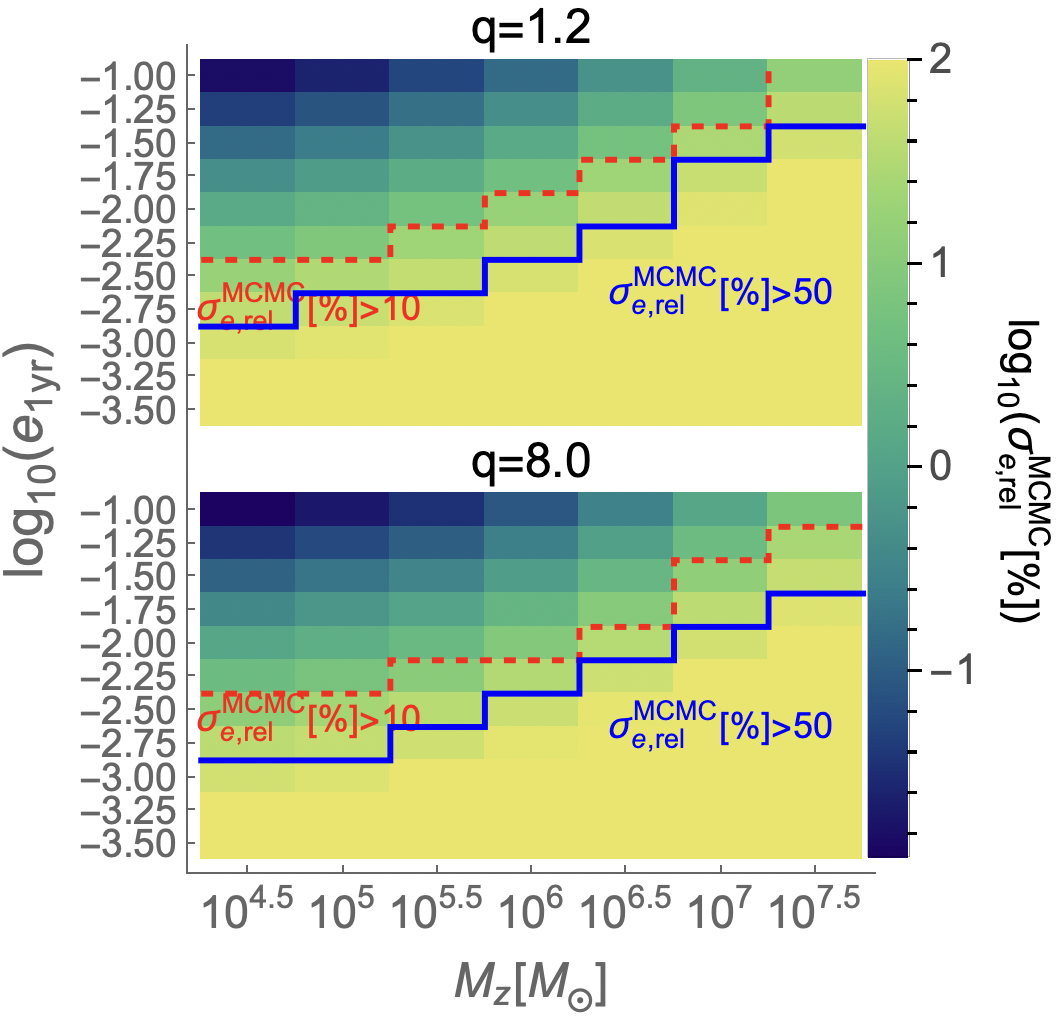}
    \caption{For the same binary parameters as in Fig.~\ref{fig:Optmatch}, relative error percentage ($\sigma^{\rm MCMC}_{e,\rm rel}[\%]$) on the recovery of eccentricity in our Bayesian inference. A dashed red line is drawn to separate the region with relative error larger than $10$ per cent and a solid blue line is drawn to separate the region with $\sigma^{\rm MCMC}_{e,\rm rel}[\%]>50$ per cent. We have suppressed relative errors above $100$ per cent to enhance the informative results.}
    \label{fig:MCMC_diff}
\end{figure}

We present the findings of our Bayesian inference in terms of $\sigma^{\rm MCMC}_{e,\rm rel}[\%]$  in Fig.~\ref{fig:MCMC_diff}. Systems with $e_{1{\rm yr}} \gtrsim 10^{-1.5}$ will mostly lead to the measurement of eccentricity to a relative error of less than $1$ per cent for lower-mass MBHBs and $\lesssim10$ per cent for higher-mass binaries, independent of $q$. The lowest value of eccentricity ($e_{\rm min}$) that LISA can measure with a less than $50$ per cent relative error is $10^{-2.75}$ for $M_z=10^{4.5}~\MSun$.

\begin{figure}
    \centering
    \includegraphics[width=0.5\textwidth]{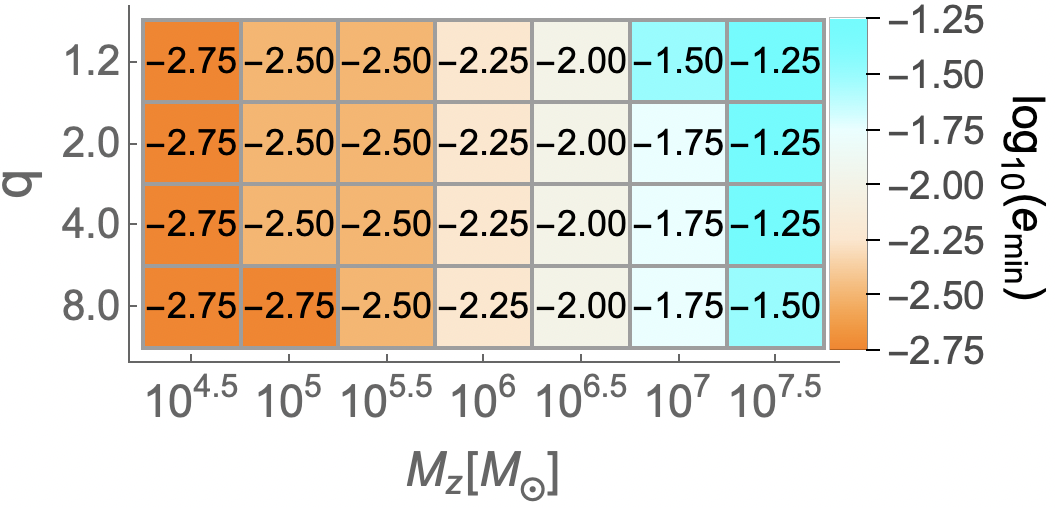}
    \caption{Minimum measurable eccentricities as a function of binary mass and mass ratio based on whether $\sigma^{\rm MCMC}_{e,\rm rel}[\%]<50$ in Eq.~\eqref{fig:min_ecc_MCMC}.}
    \label{fig:min_ecc_MCMC}
\end{figure}

We set $50$ per cent Bayesian relative error as a fiducial threshold for the measurement of eccentricity\footnote{See Appendix \ref{AppB} for the minimum eccentricity based upon the Bayes factor.}. We summarize the results of all our MCMC runs in terms of the minimum measurable eccentricity ($e_{\rm min}$) by LISA as a function of total mass and mass ratio in Fig.~\ref{fig:min_ecc_MCMC}. The results are mostly independent of mass ratio, although we witness some slight change for higher-mass ratios ($q=8$). $e_{\rm min}$ for heavier systems is around $10^{-1.5}$, whereas for lighter MBHBs the eccentricity can be measured down to $\sim10^{-2.75}$. The measurement of eccentricity in this regime can have far-reaching astrophysical consequences which we present in the discussion.

\section{Discussion}\label{Sec:discussion}

The current detectability analysis of GWs from MBHBs mostly assumes negligible eccentricity ($\lesssim10^{-4}$) once the binaries enter the LISA frequency band. However, we know that environmental interaction is necessary for binaries to reach the near-coalescence phase within a Hubble time. Therefore, it is important to consider if even residual eccentricities are measurable, which can be a tracer of the binary's environment. In this work, we remain agnostic about the driver of the binary's eccentricity. Instead, we have determined the minimum measurable eccentricity for a range of binary parameters. These limits can be compared with theoretical models of binary evolution in order to determine which binary formation scenarios lead to measurable eccentric signatures in the GW waveform. For example, we can compare our results with eccentricities predicted by binary evolution in circumbinary discs \citep{Zrake2021,DOrazio2021,Siwek2023}, which predict $e_{1\rm yr}\sim10^{-3}$ for $\sim10^{3}$--$10^{5}~\MSun$ systems at $z=1$. Based on our results in Fig.~\ref{fig:min_ecc_MCMC}, $e\sim10^{-3}$ will be indeed detectable\footnote{See Fig.~\ref{fig:MCMC2} for $e_{1\rm yr}=10^{-2.75}\approx 2\times10^{-3}$ posteriors.} for binaries within the mass range $\sim10^{4.5}$--$10^{5.5}~\MSun$ at $z=1$. Considering that the eccentricity evolution will depend on the accretion disc properties \citep{DOrazio2021}, precise detection of eccentricity in GWs can help constrain the source's environmental properties. The interaction with stars can also excite non-negligible eccentricities in the LISA band. \citet{Gualandris2022} suggest $e_{1 \rm yr}\sim10^{-4}$--$10^{-3}$ for a $4\times10^6~\MSun$ MBHB, a range of eccentricities not detectable for such massive system as per Fig.~\ref{fig:min_ecc_MCMC}. However, lower-mass binaries are not yet explored in these models. It is possible that a better waveform model which includes more physics concerning eccentricity, such as the advance of periastron \citep[][]{Tiwari2019}, could improve eccentricity measurements, but we leave this to future work. Overall, measuring specific eccentricities predicted by various environments may help to distinguish between them. Furthermore, not including eccentricity during parameter estimation could lead to significant biases in the recovery of other binary parameters (see Appendix \ref{AppB}).

In addition to measuring orbital properties of binaries in GWs, informative measurements of environmental deviations in GW waveforms are also possible for certain systems. Suppose the influence of scattered stars, surrounding gas, or a nearby third body causes alterations in the orbital evolution (compared to the same binary in vacuum). In that case, this interaction leads to a dephasing of the detected GW signal (e.g. \citealt{Garg2022,Zwick2023}), amplitude modulation due to Doppler boosting and lensing \citep{DOrazio2020}, and can excite harmonics at higher frequencies \citep{Zwick2022}. For a complete characterisation of the binary properties in astrophysical environments, it will be necessary to consider how these deviations correlate with binary parameters. Assuming one has a robust knowledge of the range of predicted residual eccentricities in different scenarios for the background (e.g. a gaseous environment versus stellar encounters) and, simultaneously, of the expected waveform modulation due to various interactions, it becomes possible to cross-correlate these parameters to enhance the determination of environmental effects. We plan to quantify the feasibility of these measurements in future work.  

LISA and other space-based mHz GW detectors will be able to observe the coalescence of MBHBs in the mass range $10^4$--$10^8 \MSun$ across the whole sky. We expect to detect at least a few events per year, with the event rate dominated by lower-mass MBH mergers at $z\lesssim 2$ \citep{AmaroSeoane2022}. However, current predictions by both post-processing of cosmological simulations and semi-analytical models vary by orders of magnitude, as they depend on intricate details of MBH seeding mechanisms and evolution in their host galaxies (e.g. \citealt{Tremmel2018,Ricarte2018,Volonteri2020,Valiante2021,Barausse2020}). While the literature is still evolving on the expected residual eccentricity at LISA entry from different environments, being able to measure the eccentricity might add important information to place further constraints on astrophysical scenarios for binary evolution. Furthermore, irrespective of that, we need to be able to extract all the potential information from the waveform if we are going to use them for fundamental physics tests, such as excluding alternative general relativistic theories (there can be various hidden degeneracies we do not know of at the moment).

The work presented here is not devoid of certain systematics that are present in the GW waveform model that is employed. As mentioned in Section~\ref{Sec:waveform}, the GW model \textsc{TaylorF2Ecc} we use only provides eccentric phase corrections up to 3PN and at $\mathcal{O}(e^2)$, which makes it reasonable to use in the low-eccentricity regime but can still induce some inaccuracies. The higher-order eccentric corrections -- up to $\mathcal{O}(e^6)$ -- are known \citep{Tiwari2019} but are cumbersome to implement within the full Bayesian inference infrastructure, and the comparison of result for the leading order in eccentricity with respect to higher-order eccentric corrections are left for future work. Additionally, \textsc{TaylorF2Ecc} does not include the component spin effects, which can have positive and negative consequences for the measurability of eccentricity. The spin-orbit and spin-spin couplings, which enter at high PN orders in the phasing, can affect the inspiral significantly \citep{Kupi2006,Brem2013,Sobolenko2017}. However, we would like to point out that LISA will very well measure spin effects near the late inspiral-merger phase of the MBH binary's evolution, where the system will be quasicircular for the eccentricities considered here, so any possible degeneracies between spins and eccentricity will be broken. To summarize, for low values of eccentricities, one can ignore the above-mentioned GW modelling issues without drastically changing the final results. 

In this work, we only consider eccentricity corrections to phase and not to the amplitude. The eccentricity enters at $\mathcal{O}(e^2)$ in phase without having a $\mathcal{O}(e)$ term which could be more important for low eccentricities. Amplitude corrections from higher harmonics induced by eccentricity can include $\mathcal{O}(e)$ terms. Therefore, it needs to be explored how much the inclusion of amplitude corrections due to eccentricity would improve the eccentricity measurement. Lower-mass MBHBs have a large number of GW cycles in the LISA band, which magnifies the $\mathcal{O}(e^2)$ terms in the cumulative phase, thereby leading to possibly better measurement of eccentricity from phase than from the amplitude for lighter binaries. Furthermore, \citet{Moore2016} states that for the small eccentricities we consider here, eccentricity corrections to phase are more important than to the amplitude.

\section{Conclusion}\label{Sec:conclusion}

In this work, we study LISA-detectable GWs from eccentric MBHBs in vacuum to find the minimum measurable eccentricity ($e_{\rm min}$) that can be inferred from the GW waveform. We consider systems that spend at least a year before merging in the LISA frequency band at $z=1$ with total redshifted mass $M_z$ in the range $10^{4.5}$--$10^{7.5}~\MSun$, primary-to-secondary mass ratio $q$ between $1.2$ and $8$, and initial eccentricity $e_{1{\rm yr}}$ from $10^{-3.5}$ to $10^{-1}$. These MBHBs have SNR $\sim100$--$2500$ (see Fig.~\ref{fig:SNR}), allowing us to infer their binary parameters with high accuracy. To robustly estimate $e_{\rm min}$, we use the inspiral-only post-Newtonian eccentric waveform template \textsc{TaylorF2Ecc}, and consider LISA's motion in its orbit around the Sun as well as time delay interferometry to suppress the laser noise by employing the \textsc{lisabeta} software. We approach this analytically via computing matches and Fisher matrices, and numerically via Bayesian inference to find $e_{\rm min}$ for optimally chosen parameter grids in Eq.~\eqref{eq:parameters} to cover our systems of interest. We itemize our findings below.

\begin{itemize}

    \item Considering only three free binary parameters -- $M_z$, $q$, and $e_{1{\rm yr}}$ -- we find that all approaches suggest that $e_{\rm min}$ mainly depends upon $M_z$ and weakly upon $q$ (see Figs~\ref{fig:Optmatch}, \ref{fig:Fisher}, and \ref{fig:MCMC_diff}).
    
    \item The optimal match-based SNR criterion, that distinguishes eccentric and quasicircular waveforms with more than $90$ per cent confidence, suggests that $e_{\rm min}$ is $\sim10^{-2.5}$ for lower-mass MBHBs ($M_z\lesssim10^{5.5}~\MSun$ and $\sim10^{-1.5}$ for higher-mass systems (see Section~\ref{Subsec:match} and Fig.~\ref{fig:Optmatch}).
    
    \item Relative errors on the recovery of eccentricity provided by the Fisher formalism for lighter systems are $\sim0.1$ per cent for high eccentricities and $\sim1000$ per cent for low $e_{1{\rm yr}}$. For heavier MBHBs, relative errors are $\sim1$ per cent for higher eccentricities and $10^5$ per cent for lower $e_{1{\rm yr}}$ (see Section~\ref{Subsec:Fisher} and Fig.~\ref{fig:Fisher}).
    
    \item Bayesian inference can constrain $e_{1{\rm yr}}\sim10^{-1.5}$ to less than $10$ per cent relative error for most MBHBs.
    
    \item Sampling also extrinsic parameters in Table~\ref{table:parameters} does not affect the eccentricity posterior significantly (see Figs~\ref{fig:Violin_ecc} and \ref{fig:MCMC_all}).
    
    \item Assuming a Bayesian relative error of less than $50$ per cent as a threshold for $e_{\rm min}$, we find that the minimum measurable eccentricity is $e_{\rm min}=10^{-2.75}$ for $10^{4.5}~\MSun$ MBHBs, independent of the mass ratio (Fig.~\ref{fig:min_ecc_MCMC}).
    
\end{itemize}

\section*{Data availability statement}

The data underlying this article will be shared on reasonable request to the authors.

\section*{Acknowledgements}

AD, MG, and LM acknowledge support from the Swiss National Science Foundation (SNSF) under the grant 200020\_192092. ST is supported by the SNSF Ambizione Grant Number: PZ00P2-202204. We acknowledge Stanislav Babak, Pedro R. Capelo, and Jonathan Gair for insightful discussions. We also thank Riccardo Buscicchio, Daniel D'Orazio, Lorenzo Speri, and Jakob Stegmann for useful comments on the manuscript. The authors also acknowledge use of the Mathematica software \citep{Mathematica}, NumPy  \citep{harris2020array}, and inspiration drawn from the \textsc{GWFAST} package \citep{Iacovelli2022} regarding the python implementation of \textsc{TaylorF2Ecc}.  

\scalefont{0.94}
\setlength{\bibhang}{1.6em}
\setlength\labelwidth{0.0em}
\bibliographystyle{mnras}
\bibliography{EccPro}
\normalsize

\appendix
\section{Circularization test for \textsc{TaylorF2Ecc}}\label{AppA}

\begin{figure}
    \centering
    \includegraphics[width=0.5\textwidth]{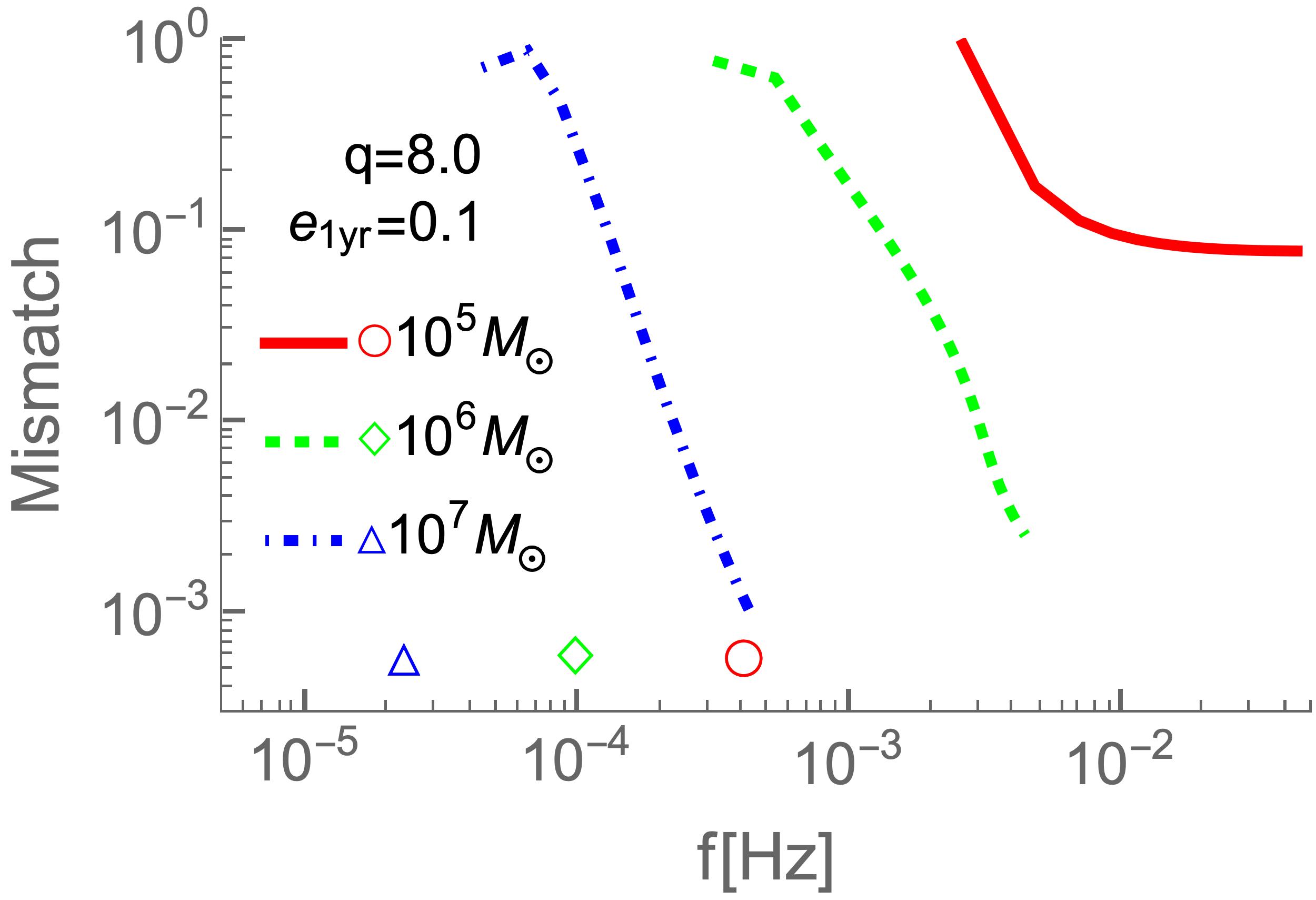}
    \caption{Tidal mismatches for three systems: $M_z=10^5~\MSun$ (solid red), $M_z=10^6~\MSun$ (dashed green), and $M_z=10^7~\MSun$ (dot-dashed blue) for $q=8.0$ and $e_{{1\rm yr}}=0.1$ over 20 frequency bins. We also indicate $f_{1\rm yr}$ for each respective binary with symbols {\Large $\circ$}, {\Large $\diamond$}, and {$\vartriangle$}.}
    \label{fig:tidalmismatch}
\end{figure}

To check that the \textsc{TaylorF2Ecc} template is well-behaved, i.e. $\tilde{h}_{\rm ecc}$ converges to $\tilde{h}_{\rm cir}$ as the system approaches the coalescence, we compute tidal matches: we divide a signal into equal frequency bins and compute the mismatch (i.e. $1-\mathscr{M}(h_{\rm cir},h_{\rm ecc})$) with respect to the cumulative frequency. In Fig.~\ref{fig:tidalmismatch}, we show the evolution of the mismatch as a function of cumulative frequency for three total masses -- $\{10^5,~10^6,~\text{and}~10^7\}~\MSun$ -- with fixed $q=8.0$ and $e_{1{\rm yr}}=0.1$ over 20 frequency bins between $f_{1\rm yr}$ and $f_{\rm ISCO}$. The figure indeed shows that the mismatch is decreasing as the MBHB approaches its ISCO, showing that \textsc{TaylorF2Ecc} is well-behaved.

\section{Bayes factor and biases}\label{AppB}
Another way to quantify whether a certain eccentricity is well recovered in our analysis is by computing the Bayes factor. For this purpose, we need to compare two hypotheses  which are trying to explain the same eccentric signal. The Null hypothesis is that a circular template (here \textsc{TaylorF2}) is enough to accurately describe this signal. The eccentric hypothesis states that you need to have the eccentricity parameter in your template (here \textsc{TaylorF2Ecc}) to properly explain this signal. We then need to take the ratio of their evidence to compute the Bayes factor
\begin{equation}
    \mathcal{B}=\frac{Z_{\rm ecc}}{Z_{\rm no~ecc}}.
\end{equation}
If $\ln\mathcal{B}>8$ \citep{Lower2018,Thrane2019}, then we have a strong evidence that the given signal comes from an eccentric system rather than a circular one. 

To do this we inject eccentric signals using \textsc{TaylorF2Ecc} and recover them with \textsc{TaylorF2} to compute $Z_{\rm no~ecc}$ and \textsc{TaylorF2Ecc} to compute $Z_{\rm ecc}$ by sampling only intrinsic parameters. Since we are in a zero noise limit and high SNR limit with only eccentricity parameter different between two models, we can compute the Savage-Dickey ratio \citep{Dickey1971} to approximately get $\mathcal{B}$. We only need to use the Fisher matrix $\Gamma_{ij}$ from \textsc{TaylorF2Ecc} for a given injected eccentricity $e_{\rm inj}$:
\begin{equation}
    \mathcal{B}\approx\pi(e)\sqrt{\frac{2\pi \bar\Gamma}{\Gamma}}\exp{\left(\frac{1}{2}\frac{e^2_{\rm inj}}{\Gamma^{-1}_{ee}}\right)},
\end{equation}
where $\Gamma$ and $\bar\Gamma$ are determinants of Fisher matrices of all parameters and parameters except eccentricity, respectively, and $\Gamma^{-1}_{ee}$ is the value of the covariance on eccentricity. Here $\pi(e)=1/(0.2-10^{-6})$ due to an uniform prior.

\begin{figure}
    \centering
    \includegraphics[width=0.5\textwidth]{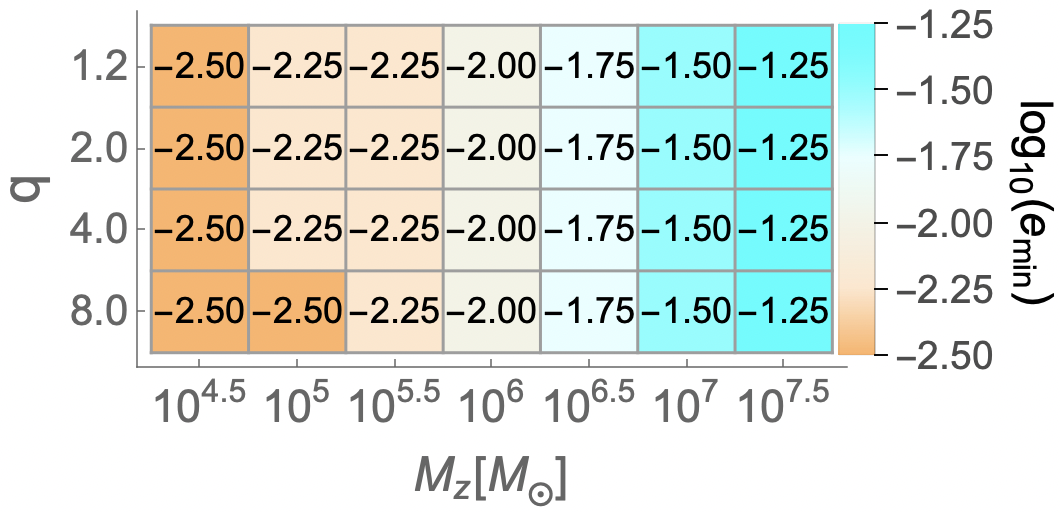}
    \caption{Same as in Fig.~\ref{fig:min_ecc_MCMC}, but now based on whether $\ln\mathcal{B}>8$.}
    \label{fig:eccmin_lnBayes}
\end{figure}

In Fig.~\ref{fig:eccmin_lnBayes}, we show minimum measurable eccentricities if $\ln\mathcal{B}>8$ for a given $M_z$ and $q$. These results are almost consistent with Fig.~\ref{fig:min_ecc_MCMC}, although 
results slightly worsen due to a stricter criterion.

We can also compute the bias induced in the estimation of $M_z$ and $q$ when fitting circular template to an eccentric signal. For this purpose, we compute the bias for a given parameter $\theta$, normalized by its standard deviation as
\begin{equation}
    \delta\theta[\sigma]=\frac{|\hat{\theta}_{\rm no~ecc}-\hat{\theta}_{\rm ecc}|}{\sigma^{\theta}_{\rm ecc}},
\end{equation}
where $\hat{\theta}$ denotes the highest likelihood point in the posterior distribution for the given model, and $\sigma^{\theta}_{\rm ecc}$ is the standard deviation of the eccentric model posterior of $\theta$. 

\begin{figure}
    \centering
    \includegraphics[width=0.5\textwidth]{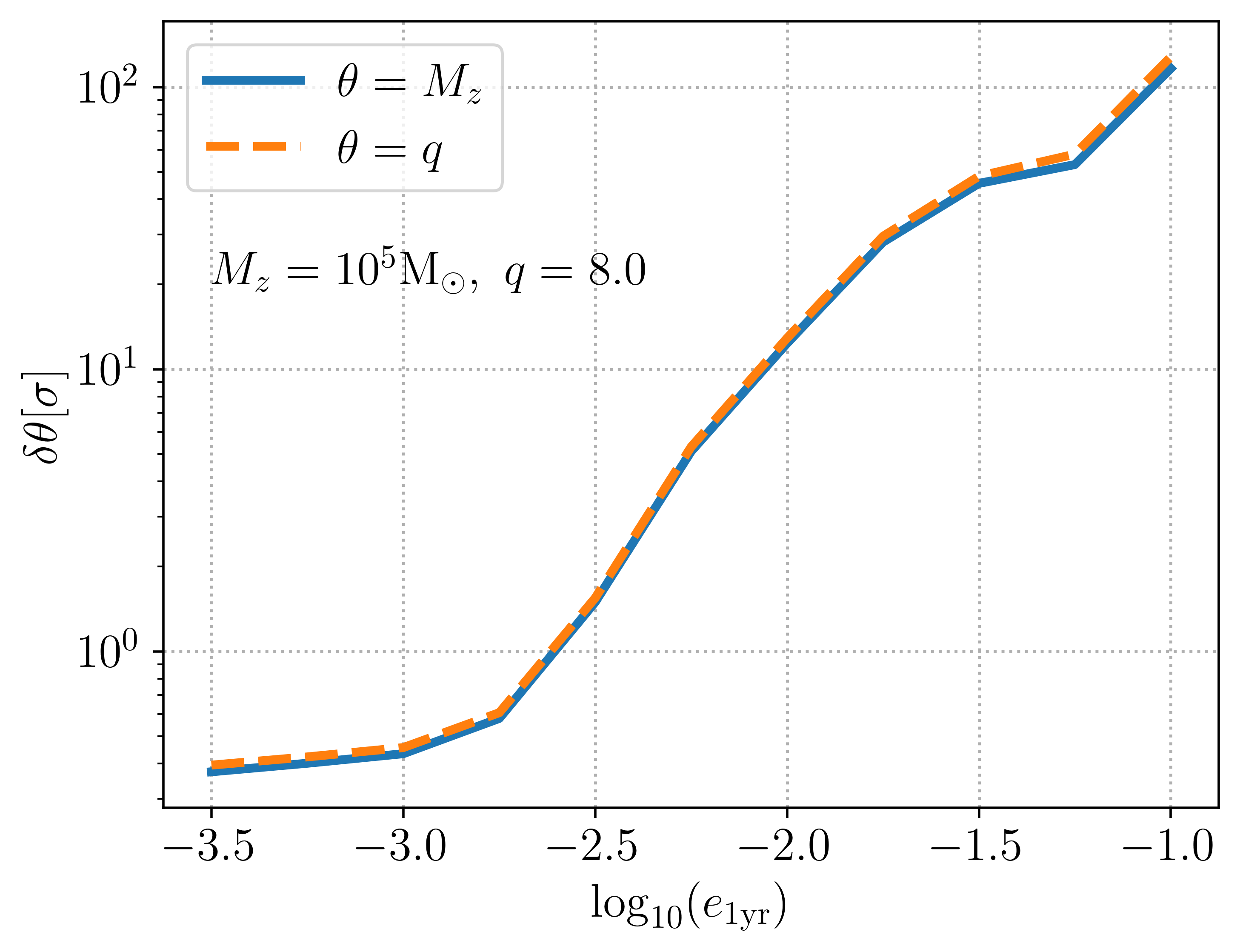}
    \caption{Biases induced in the estimation of $M_z$ and $q$ due to fitting a circular template to an eccentric signal as a function of $e_{1\rm yr}$ for a system with $M_z=10^5~\MSun$ and q=8.}
    \label{fig:Biases}
\end{figure}

In Fig.~\ref{fig:Biases}, we show $\delta\theta[\sigma]$ for $M_z$ and $q$ as a function of varying eccentricity for a fixed $M_z=10^5~\MSun$ and $q=8$. Both biases are almost identical and as expected, grow rapidly as $e_{1\rm yr}$ becomes higher. For $e_{1\rm yr}=10^{-3.5}$, the bias in both parameters is $\approx0.4$ and for $e_{1\rm yr}=0.1$, $\delta\theta[\sigma]=120$. These results emphasize the need to included eccentricity during parameter estimation.

\section{Dependence on high-frequency cutoff}\label{AppC}

\begin{figure}
    \centering
    \includegraphics[width=0.5\textwidth]{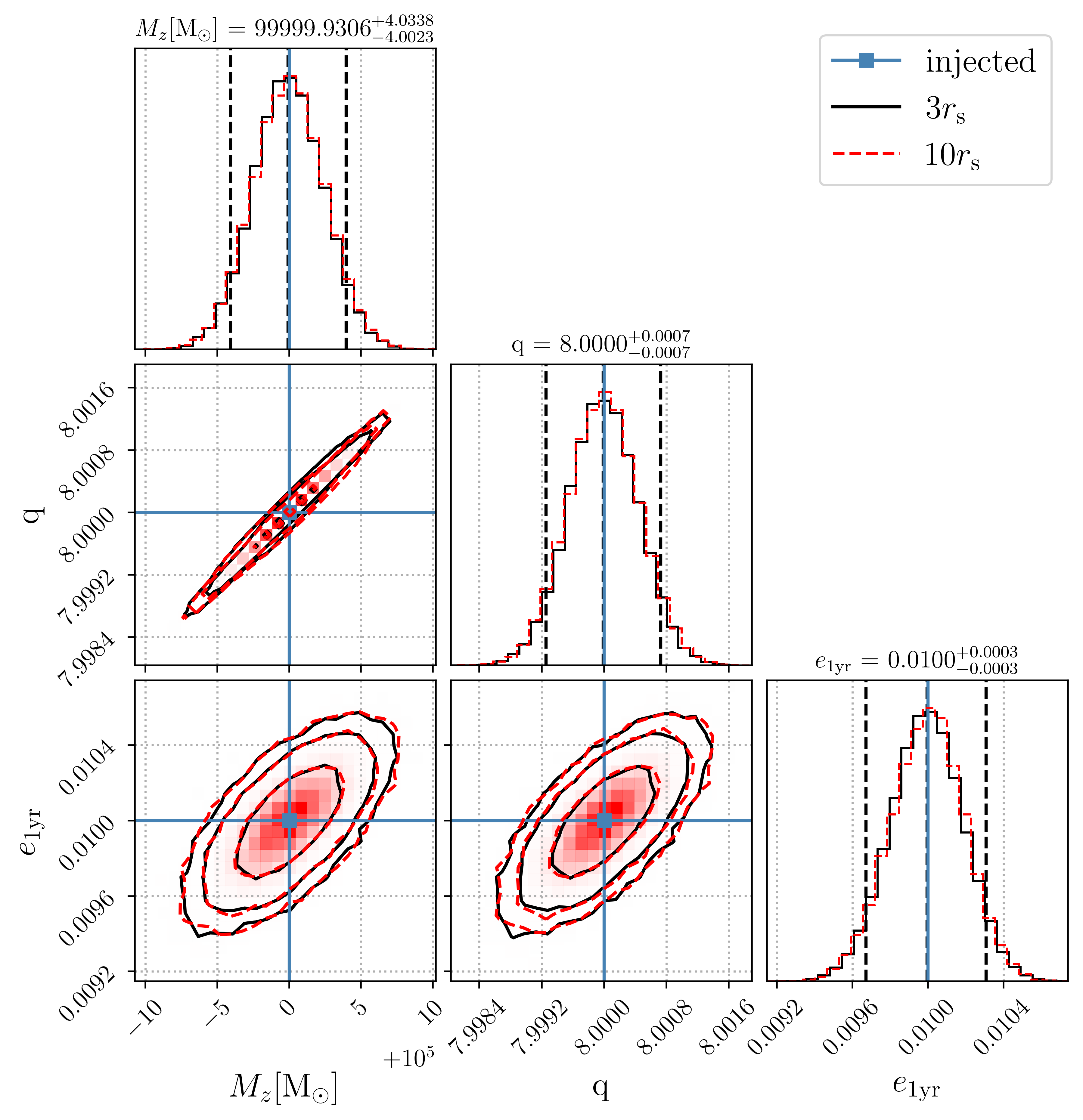}
    \caption{Posterior distributions for the same binary parameters as in Fig.~\ref{fig:MCMC} with a high frequency cutoff at $10\,r_{\rm s}$ (dashed red) binary separation compared to our fiducial cutoff at ISCO or $3\,r_{\rm s}$ separation (solid black).}
    \label{fig:MCMC_cutoff}
\end{figure}

In Fig.~\ref{fig:MCMC_cutoff}, we compare posteriors between two signals, where the high-frequency cutoff is at $10\,r_{\rm s}$ and at our fiducial cutoff $3\,r_{\rm s}$. This exercise is performed to ensure that near-merger artefacts, beyond the scope of \textsc{TaylorF2Ecc}, do not bias our results. Almost overlapping posteriors indeed illustrate that the cutoff near the merger does not affect the recovery of parameters, especially eccentricity, which is an early inspiral effect.

\section{Dependence on Fisher initialized prior's covariance}\label{AppD}

\begin{figure}
    \centering
    \includegraphics[width=0.5\textwidth]{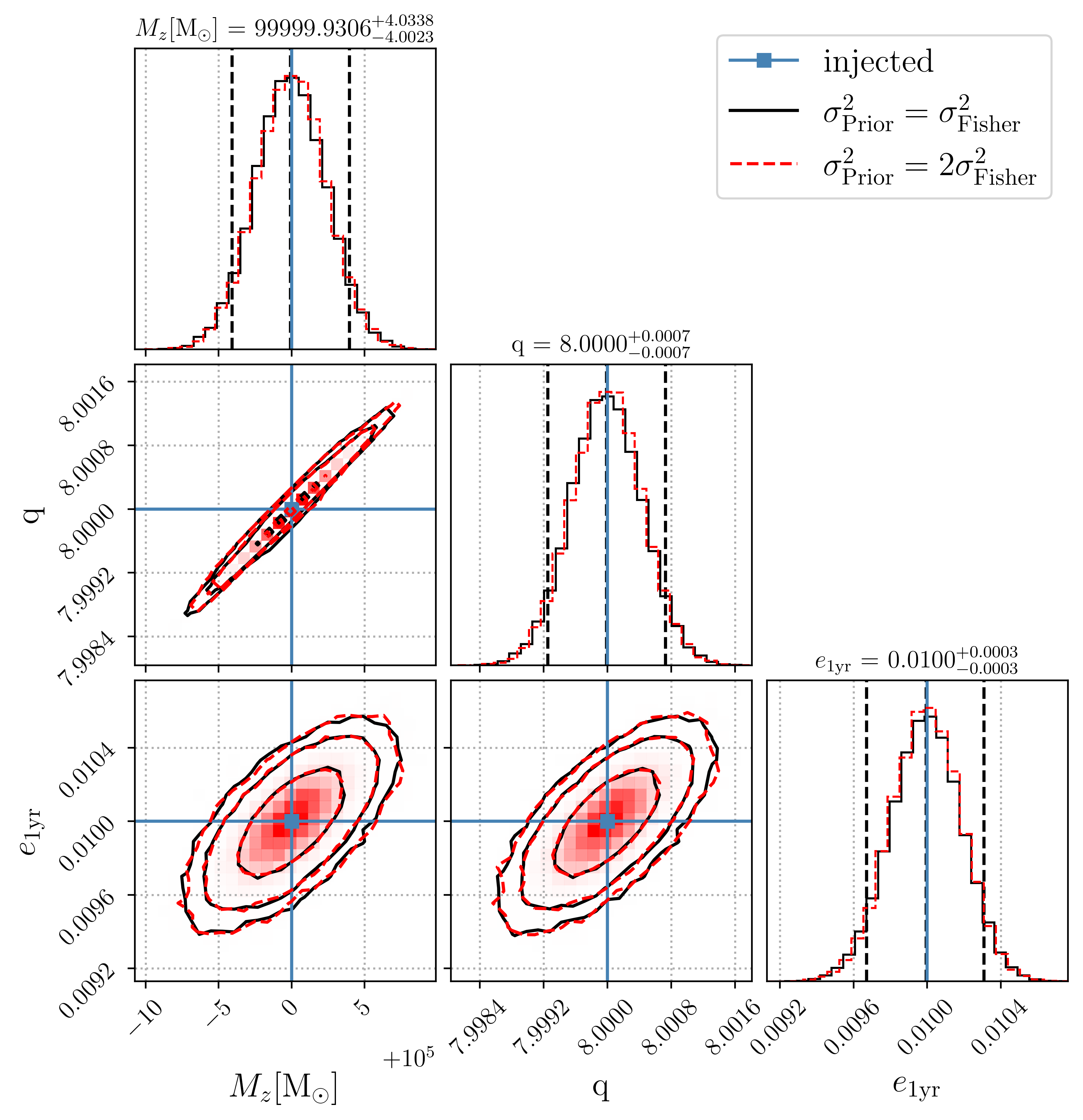}
    \caption{Posterior distributions for the same binary parameters as in Fig.~\ref{fig:MCMC} with prior's covariances ($\sigma^2_{\rm Prior}$) (solid black) being our fiducial covariances ($\sigma^2_{\rm Fisher}$) provided by the Fisher formalism compared to $\sigma^2_{\rm Prior}$ set to $2\sigma^2_{\rm Fisher}$ (dashed red).}
    \label{fig:Wider_prior}
\end{figure}

In Fig.~\ref{fig:Wider_prior}, we make a comparison between two posteriors with prior's covariances either the same as our fiducial Fisher covariances or twice the Fisher covariances. Almost identical posteriors clearly illustrate that our Bayesian runs are giving informative results and not merely giving back the Fisher priors we are using.

\section{Some interesting posteriors}\label{AppE}

\begin{figure}
    \centering
    \includegraphics[width=0.5\textwidth]{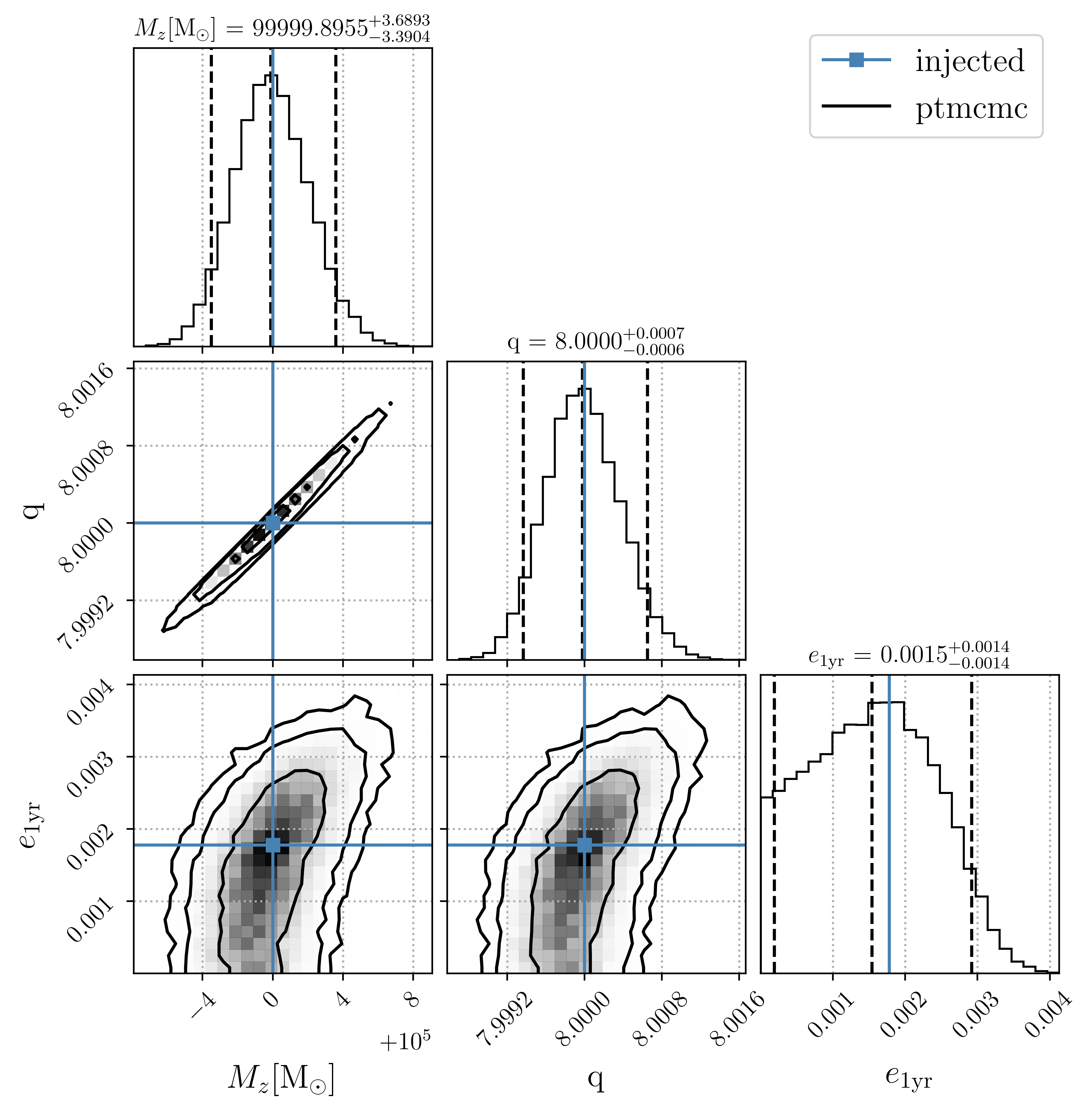}
    \caption{Same as Fig.~\ref{fig:MCMC}, but for an injected binary with $M_z=10^5~\MSun$, $q=8.0$, and $e_{1{\rm yr}}=10^{-2.75}$.}
    \label{fig:MCMC2}
\end{figure}

Fig.~\ref{fig:MCMC2} shows posteriors for intrinsic parameters for injected MBHB parameters $M_z=10^5~\MSun$, $q=8.0$, and $e_{1{\rm yr}}=10^{-2.75}$, a system that is motivated astrophysically by the binary's interaction with its environment. While the mass and the mass ratio are still well-measured as in Fig.~\ref{fig:MCMC} due to similar SNR, eccentricity posterior is broad. Nonetheless, the peak of the eccentricity posterior is very close to the injected value.

\begin{figure*}
    \centering
    \includegraphics[width=1\textwidth]{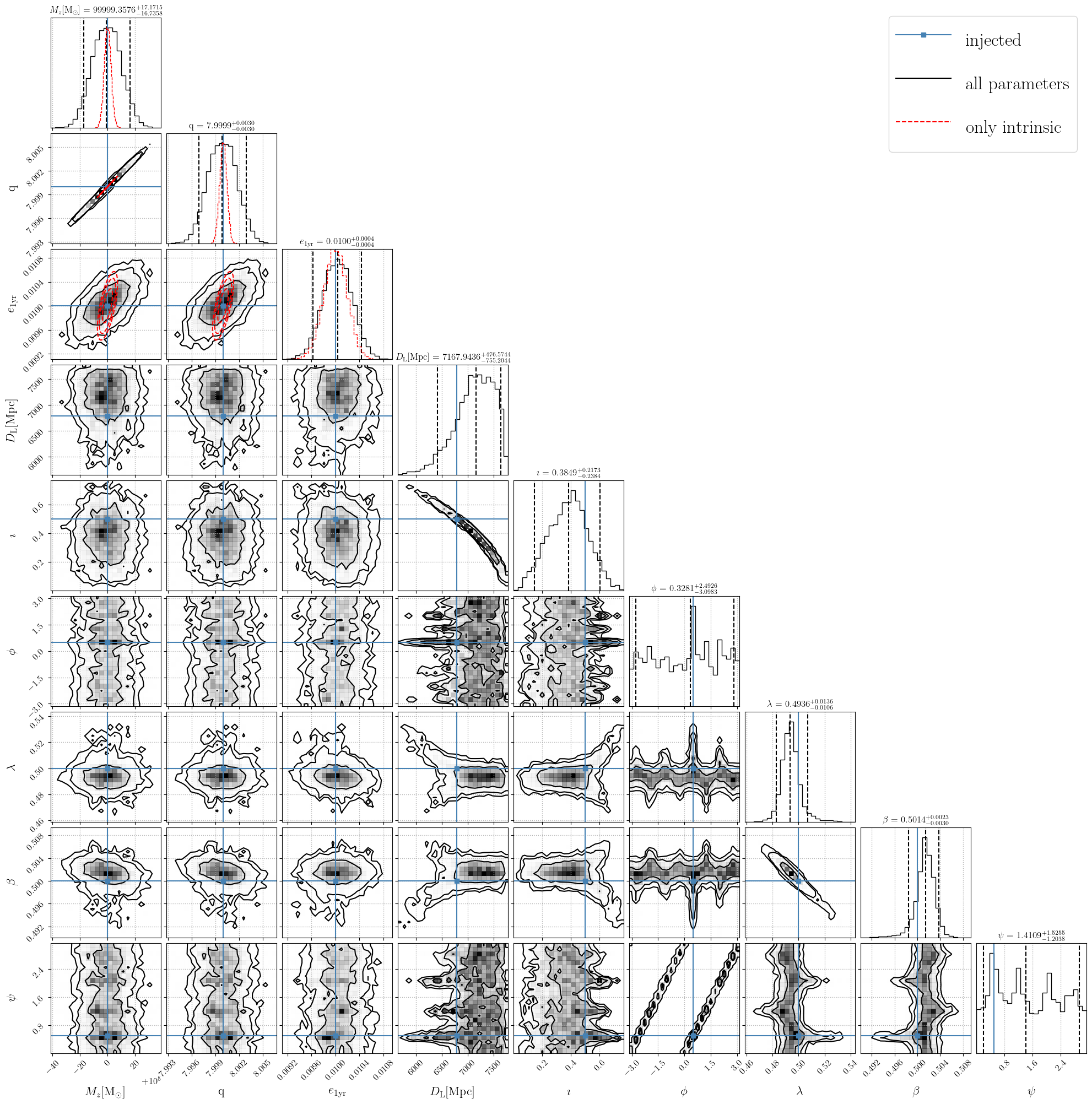}
    \caption{Posteriors (solid black) for the same intrinsic binary parameters as in Fig.~\ref{fig:MCMC} with sampling included for the extrinsic parameters. We also include posteriors (dashed red) when only sampling intrinsic parameters for comparison.}
    \label{fig:MCMC_all}
\end{figure*}

In Fig.~\ref{fig:MCMC_all}, we show posteriors for all parameters given in Table~\ref{table:parameters} for an injected binary with $M_z=10^5~\MSun$, $q=8.0$, and $e_{1{\rm yr}}=0.01$ and the extrinsic parameters set to our fiducial values. Comparison with posteriors when only varying intrinsic parameters suggest that while $M_z$ and $q$ are about order-of-magnitude less constrained, eccentricity is almost not affected due to the inclusion of extrinsic parameters, supporting inference from Fig.~\ref{fig:Violin_ecc}. As expected, there is a degeneracy between the inclination ($\imath$) and the luminosity distance ($D_{\rm L}$). The phase at coalescence ($\phi$) and the polarization angle ($\psi$) have multi-modality due to periodic functions and hence their injected values are not recovered robustly. The ecliptic latitude ($\lambda$) and longitude ($\beta$) exhibit slight degeneracies with $D_{\rm L}$ but are well constrained.

\bsp 
\label{lastpage}
\end{document}